\documentclass[draftcls,onecolumn,12pt]{IEEEtran}
\usepackage{color}
\usepackage{graphicx}
\usepackage{epstopdf}
\usepackage{amsmath}
\usepackage{amssymb}
\usepackage[english]{babel}
\usepackage{cite}
\usepackage{rotfloat}
\usepackage{mathtools}
\usepackage[justification=centering,font=small,skip=0pt]{caption}
\usepackage{amsmath}
\usepackage{makecell}
\usepackage{algorithm,algorithmic}
\usepackage{multirow}
\usepackage{subfigure}
\usepackage{booktabs}
\usepackage{colortbl}
\definecolor{kugray5}{RGB}{224,224,224}
\usepackage[normalem]{ulem}



\newcommand\cfmimo{CF mmWave massive MIMO }
\makeatletter
\newcommand{\ALOOP}[1]{\ALC@it\algorithmicloop\ #1%
	\begin{ALC@loop}}
	\newcommand{\ENDALOOP}{\end{ALC@loop}\ALC@it\algorithmicendloop}

\makeatother

\usepackage{etoolbox}
\let\mybibitem\bibitem
\renewcommand{\bibitem}[1]{%
	\ifstrequal{#1}{nature}
	{\color{blue}\mybibitem{#1}}
	{\color{black}\mybibitem{#1}}%
}

\graphicspath{ {Figures/} }

\newtheorem{theorem}{Theorem}

\newtheorem{remark}{Remark}

\newtheorem{proof}{Proof}

\newcommand{\epr}{\hfill\(\Box\)}

\DeclareCaptionLabelSeparator{periodspace}{.\quad}

\addto\captionsenglish{}
\newcommand\numberthis{\addtocounter{equation}{1}\tag{\theequation}}

\newcommand{\norm}[1]{\left\lVert#1\right\rVert} 
\newcommand{\abs}[1]{\left|#1\right|} 

\newcommand{\mean}[1]{\mathbb{E} \left\{#1\right\}}

\newcommand{\setC}{\mathbb{C}} 

\newcommand{\setA}{\mathbb{A}}

\newcommand{\mQ}{\textbf{\textit{Q}}}
\newcommand{\mR}{\textbf{\textit{R}}}
\newcommand{\mH}{\textbf{\textit{H}}} 
\newcommand{\mW}{\textbf{\textit{W}}} 

\newcommand{\mI}{\textbf{\textit{I}}}

\newcommand{\mG}{\textbf{\textit{G}}}
\newcommand{\mE}{\textbf{\textit{E}}}
\newcommand{\mF}{\textbf{\textit{F}}}

\newcommand{\mA}{\textbf{\textit{A}}}
\newcommand{\mB}{\textbf{\textit{B}}}
\newcommand{\mU}{\textbf{\textit{U}}}
\newcommand{\mV}{\textbf{\textit{V}}}

\newcommand{\mJ}{\textbf{\textit{J}}}

\newcommand{\mZ}{\textbf{\textit{Z}}}
\newcommand{\mY}{\textbf{\textit{Y}}}

\newcommand{\ve}{\textbf{\textit{e}}} 

\newcommand{\vx}{\textbf{\textit{x}}}
\newcommand{\vy}{\textbf{\textit{y}}}
\newcommand{\vr}{\textbf{\textit{r}}}

\newcommand{\vu}{\textbf{\textit{u}}}
\newcommand{\vz}{\textbf{\textit{z}}} 
\newcommand{\vh}{\textbf{\textit{h}}}

\newcommand{\vf}{\hspace{0.1cm}\textbf{\textit{f}}}
 
\newcommand{\va}{\textbf{\textit{a}}}

\newcommand{\vn}{\textbf{\textit{n}}}
\newcommand{\vnb}{\textbf{\textit{n}}^{\star}}
\newcommand{\vphi}{\boldsymbol{\varphi}}

\DeclarePairedDelimiter{\round}\lfloor\rceil

\usepackage{hyperref}  

\begin{document}
	\title{Hybrid Beamforming and Adaptive RF Chain Activation for Uplink Cell-Free Millimeter-Wave Massive MIMO Systems}
	\author{Nhan~Thanh~Nguyen,
		Kyungchun~Lee,~\IEEEmembership{Senior~Member,~IEEE}, and~Huaiyu~Dai,~\IEEEmembership{Fellow,~IEEE},
		\thanks{N. T. Nguyen and K. Lee are with the Department of Electrical and Information Engineering and the Research Center for Electrical and Information Technology, Seoul National University of Science and Technology, 232 Gongneung-ro, Nowon-gu, Seoul, 01811, Republic of Korea (e-mail: nhan.nguyen, kclee@seoultech.ac.kr).}
		\thanks{H. Dai is with the Department of Electrical and Computer Engineering, North Carolina State University, NC, USA. (e-mail: Huaiyu\_Dai@ncsu.edu).}
	}
	
	\maketitle
	\vspace{-2cm}
	\begin{abstract}
		In this work, we investigate hybrid analog--digital beamforming (HBF) architectures for uplink cell-free (CF) millimeter-wave (mmWave) massive multiple-input multiple-output (MIMO) systems. {We first propose two HBF schemes, namely, decentralized HBF (D-HBF) and semi-centralized HBF (SC-HBF). In the former, both the digital and analog beamformers are generated independently at each AP based on the local channel state information (CSI). In contrast, in the latter, only the digital beamformer is obtained locally at the AP, whereas the analog beamforming matrix is generated at the central processing unit (CPU) based on the global CSI received from all APs. We show that the analog beamformers generated in these two HBF schemes provide approximately the same achievable rates despite the lower complexity of D-HBF and its lack of CSI requirement.} Furthermore, to reduce the power consumption, we propose a novel adaptive radio frequency (RF) chain-activation (ARFA) scheme, which dynamically activates/deactivates RF chains and their connected analog-to-digital converters (ADCs) and phase shifters (PSs) at the APs based on the CSI. For the activation of RF chains, low-complexity algorithms are proposed, which can achieve significant improvement in energy efficiency (EE) with only a marginal loss in the total achievable rate.
	\end{abstract}
	
	\begin{IEEEkeywords}
		Cell-free massive MIMO, mmWave communication, hybrid beamforming, RF chain activation.
	\end{IEEEkeywords}
	\IEEEpeerreviewmaketitle
	
	\section{Introduction}

	Recently, many attempts have been made to utilize millimeter waves (mmWaves) for high-data-rate mobile broadband communications. The main challenge of mmWave communication is the large path loss due to high carrier frequencies, which significantly limits the system performance and cell coverage \cite{nguyen2018coverage, akdeniz2014millimeter}. Fortunately, the short wavelength of mmWave systems facilitates {the deployment of} massive multiple-input multiple-output (MIMO) systems, which can provide large beamforming gains to compensate for the path loss in mmWave channels. {Furthermore, Ngo \emph{et al.} in \cite{ngo2015cell, ngo2017cell} introduce a cell-free (CF) massive MIMO architecture, in which a very large number of distributed access points (APs) connected to a single central processing unit (CPU) simultaneously serve a much smaller number of users over the same time/frequency resources.} Therefore, the CF massive MIMO system is capable of providing good quality of service (QoS) uniformly to all served users, regardless of their locations in the coverage area, by using simple linear signal processing schemes \cite{ngo2015cell, ngo2017cell, femenias2019cell}. From these aspects, the combination of mmWave and CF massive MIMO systems with the deployment of large numbers of antennas {at the APs} could be a symbiotic convergence of technologies that can significantly improve the performance of next-generation wireless communication systems \cite{femenias2019cell}.
	
	\subsection{Related works}
	
	{The performance CF massive MIMO systems in conventional sub-6-GHz frequency bands have been analyzed intensively \cite{ngo2015cell, ngo2017cell, interdonato2019downlink, papazafeiropoulos2020performance, nayebi2015cell, bjornson2019making, nguyen2017energy, ngo2017total, bashar2019energy, interdonato2019scalability, bjornson2020scalable, hajri2018enhancing, hu2019cell, nayebi2017precoding, zhang2018performance, buzzi2017cell, park2018optimizing}. In particular, the closed-form expression of the achievable rate of CF massive MIMO systems is derived in \cite{ngo2015cell, ngo2017cell, interdonato2019downlink, papazafeiropoulos2020performance}. Furthermore, comparisons of CF massive MIMO and conventional small-cell MIMO systems in \cite{ngo2015cell, ngo2017cell, nayebi2015cell, bjornson2019making, papazafeiropoulos2020performance} show that the former is more robust to shadow fading correlation and significantly outperforms the latter in terms of throughput and coverage probability.} In \cite{nguyen2017energy}, a low-complexity power control technique with zero-forcing (ZF) precoding design is introduced for CF massive MIMO systems. {In \cite{ngo2017total}, an optimal power-allocation algorithm is proposed to maximize the total EE, which can double the total EE compared to the equal power control scheme. Furthermore, an AP selection (APS) scheme is proposed in \cite{ngo2017total},} in which each user chooses and connects to only a subset of APs to reduce the power consumption caused by the backhaul links. Meanwhile, in \cite{bashar2019energy}, the EE maximization problem is considered under the effect of quantization distortion of the weighted received signals at the APs. {In \cite{interdonato2019scalability, bjornson2020scalable}, the authors propose AP-clustering approaches to overcome limitations upon the scalability of CF massive MIMO systems. Specifically, the APs are grouped in clusters and multiple CPUs are used to manage these clusters, leading to a reduction in the data distribution and computational complexity involved in channel estimation, power control, and beamforming.}
	
	{Another line of work has attempted to investigate the performance of CF massive MIMO systems in mmWave channels \cite{li2018performance,alonzo2017cell,alonzo2019energy, femenias2019cell}.} In particular, Femenias et al. introduced a hybrid beamforming (HBF) framework for \cfmimo systems with limited fronthaul capacity {in} \cite{femenias2019cell}, and the eigen-beamforming scheme is applied to generate precoders/combiners. Specifically, the phases of analog beamformers are obtained by quantizing those of dominant eigenvectors of the channel covariance matrix known at the APs. In \cite{li2018performance}, a hybrid precoding algorithm leveraging antenna-array response vectors is applied to distributed MIMO systems with partially-connected  HBF architecture. Although the partially-connected HBF structure has lower power consumption than the fully-connected one, it cannot fully exploit the beamforming gains \cite{nguyen2019unequally}. In \cite{alonzo2017cell} and \cite{alonzo2019energy}, Alonzo et al. introduce an uplink multi-user estimation scheme along with low-complexity HBF architectures. Specifically, the baseband and analog precoders at each AP are generated by decomposing the fully digital ZF precoding matrix using the block-coordinate descent algorithm. {Moreover, the problems of pilot assignment and channel estimation are considered in \cite{liu2019tabu} and \cite{jin2019channel}, respectively.}
	
	In mmWave communications, a large number of antennas at the APs not only provide beamforming gains to compensate for the propagation loss in mmWave channels, but also enhance favorable propagation \cite{chen2018channel}. {In particular, it is shown in \cite{chen2018channel} that given a fixed total number of antennas at the APs, employing more antennas at fewer APs is more beneficial than deploying more APs with fewer antennas in terms of favorable propagation and channel hardening. Furthermore, large antenna arrays at the APs can achieve high beamforming gains to overcome the severe path loss in mmWave communication.  However, an excessively high power consumption is required in this deployment because signal processing in the conventional digital domain requires a dedicated RF chain and analog-to-digital converter (ADC) for each antenna. Therefore, energy-efficient HBF schemes dedicated to \cfmimo systems are required, but only limited works exist in the literature focusing on the optimization of HBF for \cfmimo systems.} Specifically, in \cite{li2018performance, alonzo2017cell, femenias2019cell, alonzo2019energy}, the analog beamformers are all separately generated at the APs based on the local CSI, which are similar to HBF schemes for small-cell mmWave massive MIMO systems when each AP in the \cfmimo system is considered as a base station in the small-cell system. {The antenna-selection (AS) schemes, which are proposed for the transceiver with a limited number of RF chains in \cite{tai2015low, liu2016energy}, can be applied to \cfmimo systems to reduce power consumption. However, they can cause performance degradation, especially for HBF in the highly correlated channels of mmWave communication \cite{gao2018low}.}
	
	\vspace{-0.5cm}
	\subsection{Contributions}
	
	
	In this work, we investigate the HBF for uplink \cfmimo systems in two scenarios: the global CSI of all APs is available or unavailable at the CPU. Then, we propose an adaptive RF chain-activation (ARFA) scheme, which provides considerable power reduction while nearly maintaining the system's total achievable rate; thus, the EE is remarkably improved. Our specific contributions can be summarized as follows:
	
	\begin{itemize}
		\item {We first propose the decentralized HBF (D-HBF) and semi-centralized HBF (SC-HBF) schemes. Both have digital beamformers generated at the AP, but their difference lies in the analog beamformer. Specifically, in SC-HBF, the analog beamforming matrices for all APs are generated at the CPU based on the global CSI. In contrast, that of the D-HBF is obtained at each AP based only on the local CSI.} {By exploiting the global CSI to jointly optimize the analog combiners at the CPU, {SC-HBF} is expected to outperform {D-HBF}. However, our analytical and numerical results show that {D-HBF} can perform approximately the same as {SC-HBF} while requiring substantially lower computational complexity and no global CSI.} 
		
		\item In \cfmimo systems with $L$ APs and $N$ RF chains at each AP, the power consumption is approximately proportional to $LN$. Because $L$ is large in CF massive MIMO systems, the total power consumption can be excessively high. To overcome this challenge, we propose an ARFA scheme. In this scheme, the RF chains are selectively activated at the APs based on {partial} CSI, and the number of active RF chains at the APs is optimized so that the proposed scheme can significantly reduce the total power consumption while causing only marginal performance loss. Our numerical analysis reveals that in \cfmimo systems, the proposed ARFA scheme with a relatively small number of active RF chains can exhibit performance comparable to that of {the conventional fixed-activation HBF scheme}, which activates all the available RF chains. As a result, a considerable improvement in the EE is achieved.

		\item In the proposed ARFA scheme, high computational complexity is required to find the optimal numbers of active RF chains at numerous APs with an exhaustive search. To reduce complexity, we propose a low-complexity near-optimal algorithms for the ARFA with {SC-HBF}. {Furthermore, ARFA is incorporated with {D-HBF} in the proposed {D-ARFA} schemes, creating a singular value-based and path loss-based {D-ARFA,} wherein the CPU requires a very limited amount of information from the APs.} Our simulation results show that the proposed algorithms perform very close to the conventional HBF scheme, in which all the available RF chains are turned on.
		
	\end{itemize}
	
	We note that an RF chain-selection (RFS) scheme is introduced in \cite{kaushik2019dynamic} for the conventional small-cell mmWave massive MIMO system. The RFS scheme and our proposed ARFA scheme are similar in exploiting a reduced number of RF chains for power reduction. However, the contributions in our work are novel in the following aspects. First, the number of active RF chains at the base station in the conventional small-cell system, which is a single integer, is optimized in the RFS scheme by solving the EE-optimization problem \cite{kaushik2019dynamic}. In contrast, we consider the \cfmimo system, and the numbers of active RF chains at numerous APs are jointly optimized by maximizing the system's total achievable throughput. This results in unequal numbers of active RF chains at the APs, and an AP can even deactivate all RF chains, leading to significant power reduction. Second, \cite{kaushik2019dynamic} focuses on optimizing the number of RF chains at the transmitter rather than at the receiver to avoid the nontrivial integer programming problem. Our work fills in this hole by optimizing the numbers of active RF chains at the receivers, which are the APs in the uplink of \cfmimo systems. Because of these systematic differences, the algorithm presented in \cite{kaushik2019dynamic} cannot be leveraged for our proposed ARFA scheme, which thus requires novel algorithms as presented in Section IV.

	The remainder of this paper is organized as follows: Section \ref{sec_model} introduces the system and channel models, whereas Section \ref{sec_HBF} describes the {D-HBF and SC-HBF} schemes. In Section \ref{sec_ARFA}, low-complexity ARFA algorithms are presented, and the power consumption of the proposed ARFA scheme is analyzed in Section \ref{sec_power}. Section \ref{sec_sim_result} presents simulation results, and the conclusion follows in Section \ref{sec_conclusion}.

	\section{System Model}
	\label{sec_model}

	We consider the uplink of a \cfmimo system, where $L$ APs, each equipped with $N_r$ receive antennas and $N (\leq N_r)$ RF chains, and $K$ single-antenna user equipments (UEs) are distributed in a large area. All APs are connected simultaneously to a CPU via fronthaul links and jointly serve $K$ UEs. At each AP, a fully connected architecture is considered for analog combining, in which $N$ RF chains are connected to $N_r$ receive antennas via a network of $NN_r$ PSs. We adopt a narrowband block-fading channel model \cite{alkhateeb2014channel, el2014spatially}. 
	
	Let $\vh_{kl}  \in \setC^{N_r \times 1}$ denote the channel between the $k$th UE and $l$th AP. In mmWave systems, $\vh_{kl}$ follows the geometric Saleh--Valenzuela channel model and is given by \cite{han2015large, lee2015hybrid, alkhateeb2014channel}
	\begin{align*}
		\vh_{kl} = \sqrt{\zeta_{kl}} \sum_{p=1}^{P_{kl}} \alpha_{kl}^{(p)}  \va_{r} (\phi_{kl}^{(p)}), \numberthis \label{eq_h}
	\end{align*}	
	where $\zeta_{kl} = \frac{G_{\mathrm{a}}}{\beta_{kl}} \frac{N_r}{P_{kl}}$. Here, $G_{\mathrm{a}}$ is the antenna gain, and $\beta_{kl}$ represents the path loss between the $k$th UE and $l$th AP, given by \cite{rappaport2015wideband, rappaport201238}
	\begin{align}
	\label{pathloss}
	\beta_{kl} \text{[dB]} = \beta_0  + 10 \epsilon \log_{10} \left(\frac{d_{kl}}{d_0}\right) + A_{\xi},
	\end{align}
	where $\beta_0  = 10 \log_{10} \left(\frac{4 \pi d_0}{\lambda}\right)^2$, $d_0 = 1$ m, $d_{kl}$ is the distance between the $k$th UE and $l$th AP, $\epsilon$ is the average path-loss exponent over distance, and $A_{\xi} \sim \mathcal{N}(0,\xi^2) $ represents the effect of shadow fading. Furthermore, $P_{kl}$ is the number of effective channel paths; $\alpha_{kl}^{(p)} \sim \mathcal{CN}(0,1), \forall l,k$ is the gain of the $p$th path;  and $\phi_{kl}^{(p)}$ is the azimuth angle of arrival (AoA). In \eqref{eq_h}, $\va_{r}(\cdot)$ represents the normalized receive array response vector at an AP, which depends on the structure of the antenna array. In this work, we consider a uniform linear array (ULA), where $\va_{r}(\cdot)$ is given by $\va_r (\phi) = \frac{1}{\sqrt{N_r}} [1, e^{j \frac{2\pi}{\lambda} d_s \sin(\phi)}, \ldots, e^{j (N_r-1) \frac{2\pi}{\lambda} d_s \sin(\phi)}]^T$
	in which $\lambda$ denotes the wavelength of the signal and $d_s$ is the antenna spacing \cite{alkhateeb2014channel}. Let $\mA_{kl} = [\va_{r} (\phi_{kl}^{(1)}), \ldots, \va_{r} (\phi_{kl}^{(P_{kl})})] \in \setC^{N_r \times P_{kl}}$ and $\boldsymbol{\alpha}_{kl} = [\alpha_{kl}^{(1)}, \ldots, \alpha_{kl}^{(P_{kl})}] \in \setC^{P_{kl} \times 1}$. Then, $\vh_{kl}$ can be equivalently given as $\vh_{kl} = \sqrt{\zeta_{kl}} \mA_{kl} \boldsymbol{\alpha}_{kl}$; thus, $\vh_{kl} \sim \mathcal{CN}(\mathbf{0}, \zeta_{kl} \boldsymbol{\Psi}_{kl})$, where $\boldsymbol{\Psi}_{kl} = \mean{\mA_{kl} \mA^H_{kl}}$.

	\subsection{Uplink Channel Estimation}	
	For channel estimation, all $K$ UEs simultaneously transmit their pilot sequences to the APs. Let $\sqrt{\tau_p} \vphi_k \in \setC^{\tau_p \times 1}$ be the pilot sequence of the $k$th UE, where $\norm{\vphi_k}^2 = 1$, $k=1,\ldots,K$. Here, $\tau_p$ $(\tau_p < \tau_c)$ is the length of $\vphi_k$, where $\tau_c$ denotes the length of each coherence interval (in samples). When all $K$ UEs send their pilots, the received signal at the $l$th AP is given as:
	\begin{align*}
		\mY_{l} = \sqrt{\tau_p \rho_p} \sum_{k=1}^K \vh_{kl} \vphi_k^H + \mZ_{l}, \numberthis \label{eq_Y_train}
	\end{align*}
	where $\rho_p > 0$ represents the average transmit power of each UE; $\mZ_{l} \in \setC^{N_r \times \tau_p}$ is the receiver noise, whose entries are independent and identically distributed (i.i.d.) $\mathcal{CN}(0,\sigma^2)$ random variables (RVs); and $\sigma^2$ is the noise power. To estimate $\vh_{kl}$, $\mY_{l}$ is projected onto $\vphi_k$, which yields
	\begin{align*}
		\vy_{kl} \triangleq \mY_{l} \vphi_k = \sqrt{\tau_p \rho_p} \vh_{kl} + \sqrt{\tau_p \rho_p} \sum_{i \neq k}^K \vh_{il} \vphi_i^H \vphi_k + \mZ_{l} \vphi_k.
	\end{align*}
	Thus, the minimum mean-square error (MMSE) estimate of $\vh_{kl}$, denoted by $\hat{\vh}_{kl}$, is given by
	\begin{align*}
		\hat{\vh}_{kl} = \mean{\vh_{kl} \vy_{kl}^H} \left(\mean{\vy_{kl} \vy_{kl}^H}\right)^{-1} \vy_{kl}  = \sqrt{\tau_p \rho_p} \zeta_{kl}  \boldsymbol{\Psi}_{kl}  \left(\tau_p \rho_p  \sum_{i =1}^K  \zeta_{il}  \boldsymbol{\Psi}_{il} \abs{\vphi_i^H \vphi_k}^2 + \sigma^2 \mI_{N_r} \right)^{-1} \vy_{kl}. \numberthis \label{eq_estimation}
	\end{align*}

	Assume that knowledge of the correlation matrix $\boldsymbol{\Psi}_{kl}$, $\forall k$, is available at the $l$th AP \cite{bjornson2019making}, from which $\hat{\vh}_{kl}$ can be determined. Note that in (4), we assume that the signals received at all the antennas of the AP are available for MMSE estimation. As a result, the estimate of the full CSI associated with all the antennas can be obtained via the low-complexity MMSE estimator. In a very slow-fading and sparse channel, the full CSI can be obtained by a compressed sensing-based approach \cite{alkhateeb2014channel}, but with the high complexity required for a large number of compressed-sensing measurements \cite{alkhateeb2015compressed}.
	
	Let $\hat{\mH}_{l} \triangleq [ \hat{\vh}_{1l}, \ldots, \hat{\vh}_{Kl} ] \in \setC^{N_r \times K}$ denote the estimated channel matrix between the $K$ UEs and $l$th AP. Furthermore, we define $\hat{\mH} \triangleq \left[ \hat{\mH}_{1}^T, \ldots, \hat{\mH}_{L}^T \right]^T \in \setC^{LN_r \times K}$ as the composite estimated channels between all the APs and UEs. In the next section, $\hat{\mH}_{l}$ and $\hat{\mH}$ are employed for the hybrid beamforming design of the uplink data transmission.
	
	\subsection{Uplink Data Transmission}
	Denote by $x_k$ the symbol sent from the $k$th UE to all the APs, such that $\mathbb{E} \{\abs{x_k}^2\} = 1$, $\forall k$. The signal input-output relationship at the $l$th AP can be expressed as
	\begin{align*}
		\vr_l = \sqrt{\rho} \sum_{k=1}^{K} \mW_l^H \mF_l^H \vh_{kl} x_k + \mW_l^H \mF_l^H \vz_l, \numberthis \label{eq_vr_l}
	\end{align*}	
	where $\rho$ represents the average transmit power, and $\vz_l$ is the noise vector, whose elements are i.i.d. $\mathcal{CN}(0,\sigma^2)$ RVs. Furthermore, $\mF_l \in \setC^{N_r \times N}$ is the analog combining matrix at the $l$th AP. Its $n$th column, i.e., $\vf_{ln} = [f_{ln}^{(1)}, \ldots, f_{ln}^{(N_r)}]^T$, is the analog weight vector corresponding to the $n$th RF chain at the $l$th AP, and $f_{ln}^{(i)} = \frac{1}{\sqrt{N_r}} e^{j \theta_{ln}^{(i)}}$ is the $i$th element of $\vf_{ln}$. $\mW_l \in \setC^{N \times K}$ denote the digital combining matrix at the $l$th AP. Then, the APs send $\vr_l, \forall l$ to the CPU via a fronthaul network to perform the final signal detection. In this work, we assume a simple centralized decoding scheme at the CPU, which requires minimal information exchange between the APs and CPU. In this scheme, the final decoded signal at the CPU is given as the average of the local estimates, that is, $\frac{1}{L} \sum_{l=1}^{L} \vr_l$ \cite{bjornson2019making}. 
	
	The composite received signal available at the CPU can be expressed as
	\begin{align}
	\begin{bmatrix} \vr_1 \\ \vdots \\ \vr_L \end{bmatrix}
	= \sqrt{\rho} \sum_{k=1}^{K} 
	\begin{bmatrix} \mW_1^H \mF_1^H \vh_{k1} \\ \vdots \\ \mW_L^H \mF_L^H \vh_{kL} \end{bmatrix}
	x_k
	+ 
	\begin{bmatrix} \mW_1^H \mF_1^H \vz_1 \\ \vdots \\ \mW_L^H \mF_L^H \vz_L \end{bmatrix}. \label{system_model_2}
	\end{align}
	Let $\mF \triangleq \text{diag} \left\{ \mF_1, \ldots, \mF_L \right\} \in \setC^{LN_r \times LN}$ and $\mW \triangleq \text{diag} \left\{ \mW_1, \ldots, \mW_L \right\}   \in \setC^{LN \times LK}$ be block-diagonal matrices containing the analog and digital combiners for all $L$ APs. In this work, we refer to $\mF$ and $\mW$ as \textit{global combiners}, whereas $\left\{ \mF_1, \ldots, \mF_L \right\}$ and $\left\{ \mW_1, \ldots, \mW_L \right\}$ for the signal combined at APs $\{1,\ldots, L\}$ are referred to as the \textit{local combiners}. Furthermore, let $\mH_{l} \triangleq \left[ \vh_{1l}, \ldots, \vh_{Kl} \right] \in \setC^{N_r \times K}$ denote the channel matrix between the $K$ UEs and the $l$th AP. We define $\vr \triangleq [\vr_1^T, \ldots, \vr_L^T]^T \in \setC^{LK \times 1}$, $\vx \triangleq [x_1, \ldots, x_K]^T \in \setC^{K \times 1}$, $\vz \triangleq [\vz_1^T, \ldots, \vz_L^T]^T \in \setC^{LN_r \times 1}$, and $\mH \triangleq [\mH_1^T, \ldots, \mH_L^T]^T \in \setC^{LN_r \times K}$.
	Then, \eqref{system_model_2} can be rewritten in a more compact form as
	\begin{align*}
		\vr &= \sqrt{\rho} \mW^H \mF^H \mH \vx  + \mW^H \mF^H \vz = \sqrt{\rho} \mW^H \mF^H \mH \vx  + \tilde{\vz}, \numberthis \label{system_model_3}
	\end{align*}
	where $\tilde{\vz} \triangleq \mW^H \mF^H \vz \sim \mathcal{CN}(0, \mR_{\tilde{z}})$, and $\mR_{\tilde{z}} \triangleq  \sigma^2 \mW^H \mF^H \mF \mW$.
	
	We note that the analog processing is separately performed at the APs because the ADCs, RF chains, and PSs are installed at the APs. However, $\left\{ \mF_1, \ldots, \mF_L \right\}$ can be generated either at the APs based on their local CSI or at the CPU based on the global CSI. Once the analog combiner is obtained, digital processing can be carried out at the corresponding AP. We follow the common assumption in \cite{ngo2017cell, buzzi2017cell} that the digital combining is performed at the APs individually. Therefore, in this work, the D-HBF scheme refers to the HBF with analog combiners generated at each AP separately, whereas SC-HBF implies that the analog combiners are computed at the CPU based on the global CSI. In this regard, we note that the perfect local/global CSI is not available at neither the APs nor the CPU, respectively. Therefore, to perform D-HBF, the AP, say the $l$th AP, treats its own estimated channels $\hat{\mH}_l$ as the true channel and employs it to generate the hybrid combiners $\mF_l$ and $\mW_l$. Similarly, in SC-HBF, the CPU exploits the global estimated CSI $\hat{\mH}$ to obtain $\left\{ \mF_1, \ldots, \mF_L \right\}$.
	
	\section{{SC-HBF and D-HBF}}
	\label{sec_HBF}
	
	Based on \eqref{system_model_3}, the total achievable rate $\bar{\mathcal{R}}$ can be expressed as \cite{el2014spatially} 
	\begin{align*}
	\bar{\mathcal{R}} = \frac{\tau_c-\tau_p}{\tau_c} \log_2 \abs{\mI_{LK} + \rho \mR_{\tilde{z}}^{-1} \mW^H \mF^H \mH \mH^H \mF \mW}. \numberthis \label{eq_rate}
	\end{align*}	
	We aim to design hybrid combiners that maximize $\mathcal{R}$. The design of $\mF$ and $\mW$ can be decoupled by first designing the analog combiner assuming an optimal digital combiner and then finding the optimal digital combiner for the derived analog one \cite{sohrabi2016hybrid}. Therefore, the analog beamforming design problem is formulated as 
	\begin{subequations}
		\begin{align*}
		(\text{P}_{\mathrm{a}}) \quad \underset{\{\mF_1, \ldots, \mF_L\}}{\textrm{maximize}} \quad &  \log_2 \abs{\mI_{LN} + \rho \left(\mF^H \mF\right)^{-1} \mF^H \mH \mH^H \mF} \numberthis \label{Pa_obj}, \\
		\textrm{subject to} \quad 
		& \mF = \text {diag} \left\{ \mF_1, \ldots, \mF_L \right\}  \numberthis. \label{Pa_cons_1}\\
		&\vf_{ln} \in \mathcal{F}, \forall l,n, \numberthis \label{Pa_cons_2}
		\end{align*}
	\end{subequations}
	where $\gamma = \frac{\rho}{\sigma^2}$, and $\mathcal{F}$ is the set of feasible analog combining coefficients $f_{ln}^{(i)} = \frac{1}{\sqrt{N_r}} e^{j \theta_{ln}^{(i)}}, \forall l,n, i$. To simplify the objective function in $(\text{P}_{\mathrm{a}})$, we assume $\mF^H_l \mF_l \approx \mI_{N}$ \cite{el2014spatially, sohrabi2016hybrid}, which is tight in the considered \cfmimo system with a sufficiently large number of antennas deployed at each AP. Consequently, we have $\mF^H \mF \approx \mI_{LN}$, and the objective function in \eqref{Pa_obj}, which is the sum rate achieved by analog combining, can be approximated by $\frac{\tau_c-\tau_p}{\tau_c} \log_2 \abs{\mI_{LN} + \gamma \mF^H \mH \mH^H \mF} \triangleq \bar{\mathcal{R}}^{\mathrm{a}}$. Therefore, the optimal analog combiners can be solved approximately in
	\begin{align*}
	(\text{P}'_{\mathrm{a}}) \quad \underset{\{\mF_1, \ldots, \mF_L\}}{\textrm{maximize}} \quad   &\bar{\mathcal{R}}^{\mathrm{a}} \\
	\textrm{subject to} \quad &\eqref{Pa_cons_1}, \eqref{Pa_cons_2}. \numberthis
	\end{align*}
	The objective function $\bar{\mathcal{R}}^{\mathrm{a}}$ of $(\text{P}'_{\mathrm{a}})$ is further investigated in the following theorem.
	
	{\begin{theorem}
			\label{theorem_rate_C}
			In a \cfmimo system with $L$ APs, we have $\bar{\mathcal{R}}^{\mathrm{a}} =  \sum_{l=1}^{L} \bar{\mathcal{R}}^{\mathrm{a}}_l$, where
			\begin{align*}
			\bar{\mathcal{R}}^{\mathrm{a}}_l = \frac{\tau_c-\tau_p}{\tau_c} \log_2 \det \left( \mI_{N} + \gamma \mF_l^H \mH_l \mQ_{l-1}^{-1} \mH_l^H \mF_l \right), \numberthis \label{R_n}
			\end{align*}
			with $\mQ_{0} = \mI_{K}$ and
			\begin{align*}
			\mQ_{l-1}  = \mQ_{l-2} + \gamma \mH_{l-1}^H \mF_{l-1} \mF_{l-1}^H \mH_{l-1}. \numberthis \label{Q_n_1}
			\end{align*}
	\end{theorem}}
	
	\begin{proof}
		See Appendix \ref{appendix_rate_C}.\epr
	\end{proof}
	
	Based on Theorem \ref{theorem_rate_C}, $\bar{\mathcal{R}}^{\mathrm{a}}$ can be maximized by optimizing $\{\bar{\mathcal{R}}^{\mathrm{a}}_1, \ldots, \bar{\mathcal{R}}^{\mathrm{a}}_L\}$ corresponding to APs $\{1, \ldots, L\}$. We recall that the $l$th AP treats $\hat{\mH}_l$ as the true channel to obtain the combiners. Therefore, $\mF_l^{\star}$ in $(\text{P}'_{\mathrm{a}})$ can be determined by maximizing $\mathcal{R}^{\mathrm{a}}_l \triangleq \frac{\tau_c-\tau_p}{\tau_c} \log_2 \det \left( \mI_{N} + \gamma \mF_l^H \hat{\mH}_l \hat{\mQ}_{l-1}^{-1} \hat{\mH}_l^H \mF_l \right)$, where $\hat{\mQ}_l$ is defined similarly to $\mQ_l$ by replacing $\mH_l$ with $\hat{\mH}_l$. Consequently, we can write
	\begin{align*}
	\mF_l^{\star} = \arg \max_{\mF_l} \mathcal{R}^{\mathrm{a}}_l, \forall l, \hspace{0.15cm} \text{subject to} \hspace{0.1cm} \vf_{l1}, \ldots, \vf_{lN} \in \mathcal{F}. \numberthis \label{opt_2}
	\end{align*}
	Let $\left\{ {\vu}^{\star}_{l1}, \ldots, {\vu}^{\star}_{lN} \right\}$ be the $N$ singular vectors corresponding to $N$ largest singular values of $\hat{\mH}_l \hat{\mQ}_{l-1}^{-1} \hat{\mH}_l^H$, which are in decreasing order. {Then, columns $\{\vf_{l1}^{\star}, \ldots, \vf_{lN}^{\star}\}$ of a near-optimal solution to \eqref{opt_2} can be obtained by quantizing $\left\{ {\vu}^{\star}_{l1}, \ldots, {\vu}^{\star}_{lN} \right\}$, respectively, to the nearest vector in $\mathcal{F}$ \cite{li2018performance}, i.e.,}
	\begin{align*}
	\vf_{ln}^{\star} = \arg \min_{\vf_{ln} \in \mathcal{F}} \norm{\vu_{ln}^{\star} - \vf_{ln}}^2, \forall n. \numberthis \label{quan_fln}
	\end{align*}
	At the $l$th AP, once the analog combiner $\mF_l^{\star}$ is found, the optimal digital combiner is given as the MMSE solution, i.e.,
	\begin{align*}
	\mW_l^{\star} = \mJ^{-1} \mF_l^{\star H} \hat{\mH}_l, \numberthis \label{eq_WMMSE}
	\end{align*}
	where $\mJ = \mF_l^{\star H} \hat{\mH}_l \hat{\mH}_l^H \mF_l^{\star} + \frac{1}{\gamma} \mF_l^{\star H} \mF_l^{\star}$ \cite{sohrabi2016hybrid}. In the following subsections, we propose two HBF schemes in which the analog combiners are derived based on different assumptions for CSI.

	\begin{algorithm}[t]
		\small
		\caption{{SC-HBF} scheme}
		\label{algorithm_C_HBF}
		\begin{algorithmic}[1]
			\ENSURE {$\left\{\mF_1^{\star}, \ldots, \mF_{L}^{\star}\right\}$, $\left\{\mW_1^{\star}, \ldots, \mW_{L}^{\star}\right\}$, and $\left\{ \mathcal{R}^{\mathrm{a}}_1, \ldots, \mathcal{R}^{\mathrm{a}}_L \right\}$.}
			\STATE {{At the CPU: }$\hat{\mQ}_{0} = \mI_K$}
			
			\FOR {$l = 1 \rightarrow L$}

			\FOR {$n = 1 \rightarrow N$}
			\STATE {Set $\vu_{ln}^{\star}$ to the singular vector corresponding to the $n$th largest singular value of $\hat{\mH}_l \hat{\mQ}_{l-1}^{-1} \hat{\mH}_l^H$.}
			\STATE {$\vf_{ln}^{\star} = \frac{1}{\sqrt{N_r}} \mathcal{Q} (\vu_{ln}^{\star})$}
			\ENDFOR
			
			\STATE {$\mF_l^{\star} = \left[ \vf_{l1}^{\star}, \ldots, \vf_{lN}^{\star} \right]$}
			
			\STATE {$\mG_l = \hat{\mH}_l^H \mF_l^{\star} {\mF_l^{\star}}^H \hat{\mH}_l$}
			\STATE {$\hat{\mQ}_{l} = \hat{\mQ}_{l-1} + \gamma \mG_{l}$}
			\STATE $\mathcal{R}_l^{\mathrm{a}} = \log_2 \left( \mI_{N} + \gamma {\mF_l^{\star}}^H \hat{\mH}_l \hat{\mQ}_{l-1}^{-1} \hat{\mH}_l^H \mF_l^{\star} \right)$
			\ENDFOR
			
			\STATE {At the $l$th AP: compute $\mW_l^{\star}$ based on \eqref{eq_WMMSE}.}
			
		\end{algorithmic}
	\end{algorithm}
	
	\subsection{{SC-HBF}}
	
	It is evident from \eqref{R_n} and \eqref{Q_n_1} that $\mathcal{R}^{\mathrm{a}}_l$ depends not only on $\hat{\mH}_l$, but also on $\hat{\mH}_{l-1}, \hat{\mH}_{l-2}, \ldots, \hat{\mH}_1$. Therefore, finding $\mF_l^{\star}$ requires not only $\hat{\mH}_{l}$ but also $\hat{\mH}_{l-1}, \hat{\mH}_{l-2}, \ldots, \hat{\mH}_1$. This is similar to the requirements for determining analog beamformers for sub-arrays in the partially-connected HBF architecture \cite{gao2016energy, nguyen2019unequally}. As a result, solving $\left\{\mF_1^{\star}, \mF_2^{\star}, \ldots, \mF_{L}^{\star}\right\}$ requires the CSI of the channels between all $L$ APs and $K$ UEs, i.e., $\left\{\hat{\mH}_1, \hat{\mH}_2, \ldots, \hat{\mH}_L \right\}$, which can be available at the CPU; hence, finding $\mF^{\star}$ based on \eqref{opt_2} requires a {SC-HBF} scheme. 
	
	Algorithm \ref{algorithm_C_HBF} presents the proposed {SC-HBF} scheme to obtain $\left\{\mF_1^{\star}, \mF_2^{\star}, \ldots, \mF_{L}^{\star}\right\}$. In particular, in steps 3--6, the combining vector $\vf_{ln}^{\star}$ is obtained by quantizing $\vu_{ln}^{\star}$ based on \eqref{quan_fln}, which ensures that the resultant analog combiners belong to the feasible set $\mathcal{F}$. Then, $\mF_l^{\star}$ is found in step 7 and $\mG_{l}$ is computed in step 8, followed by $\hat{\mQ}_l$ being updated in step 9. In step 10, $\mathcal{R}^{\mathrm{a}}_l$ corresponding to the $l$th AP is computed. {Furthermore, the digital combiner is computed at each AP, as in step 12. We note that in Algorithm \ref{algorithm_C_HBF}, steps 1--11 are performed at the CPU, whereas step 12 is performed at the APs.}
	
	\vspace{-0.5cm}
	\subsection{{D-HBF}}
	\label{sec_SCHBF}
	Let $\left\{ \tilde{\vu}^{\star}_{l1}, \ldots, \tilde{\vu}^{\star}_{lN} \right\}$ be the $N$ singular vectors corresponding to the $N$ largest singular values of $\hat{\mH}_l$, which are in decreasing order. Furthermore, define
	\begin{align*}
	\tilde{\vf}_{ln}^{\star} = \arg \min_{\vf_{ln} \in \mathcal{F}} \norm{\tilde{\vu}_{ln}^{\star} - \vf_{ln}}^2, \forall n. \numberthis \label{f_tilde}
	\end{align*}
	Then, in the {D-HBF} scheme, the optimal local analog combiner generated at the $l$th AP based on $\hat{\mH}_l$ can be given as $\tilde{\mF}_l^{\star}= \left[ \tilde{\vf}_{l1}^{\star}, \ldots, \tilde{\vf}_{lN}^{\star} \right]$ \cite{li2018performance}. Let $\tilde{\mF}^{\star} = \text{diag} \left\{ \tilde{\mF}_1^{\star}, \ldots, \tilde{\mF}_L^{\star} \right\}$. In the following theorem, we show that the total achievable rate achieved by analog combining in the {D-HBF} scheme is approximately equal to that in {SC-HBF}.
	
	\begin{remark}
	    \label{remark_rate_SC}
		In CF mmWave massive MIMO systems with large $L$ and low SNRs due to the significant pathloss in the mmWave channels, {the total achievable rate achieved by the analog combining in the {D-HBF} scheme, denoted by $\tilde{\mathcal{R}}^{\mathrm{a}}$, is approximately the same as that of the {SC-HBF} scheme, i.e., 
			\begin{align*}
			\tilde{\mathcal{R}}^{\mathrm{a}} = \frac{\tau_c-\tau_p}{\tau_c} \log_2 \det \left(\mI_{K} +  \gamma \mH^H \tilde{\mF}^{\star} \tilde{\mF}^{{\star} H} \mH \right) \approx \bar{\mathcal{R}}^{\mathrm{a}}, \numberthis \label{R_SC}
			\end{align*}
			where $\bar{\mathcal{R}}^{\mathrm{a}}$ is given in Theorem \ref{theorem_rate_C}.}
	\end{remark}
		
	\begin{proof}
		See Appendix \ref{appendix_rate_SC}.\epr
	\end{proof}
	
	It is observed that {D-HBF} can be performed with considerably lower computational complexity than {SC-HBF}. Specifically, only $N$ singular vectors corresponding to the $N$ largest singular values of the channel matrix are required to form the analog combiner. In contrast, additional matrix inversions, multiplications, and additions are performed in steps 4, 8, and 9 of Algorithm \ref{algorithm_C_HBF} for the {SC-HBF} scheme. Notably, despite the simpler implementation and lower complexity of the {D-HBF} scheme, it can approximately achieve the performance of {SC-HBF}, as stated in Remark \ref{remark_rate_SC}. {Furthermore, the D-HBF scheme requires less information exchange between the APs and CPU. Specifically, only $K $ complex numbers in $\vr_l$ are sent to the CPU on the fronthaul link to perform the final soft detection \cite{ngo2017cell, buzzi2017cell}, whereas the exchange of the CSI and analog combining matrix is not required, in contrast to the SC-HBF scheme}
	
	Remark \ref{remark_rate_SC} indicates that an efficient analog beamforming matrix can be designed based on the local CSI available at each AP of \cfmimo systems. However, this does not mean that further information exchange via the fronthaul links is completely useless. Indeed, the information exchange between the APs and CPU in CF massive MIMO systems can also be exploited to improve the EE. In the next section, it is discussed that by adaptively activating RF chains based on global CSI in {SC-HBF} or on limited information in {D-HBF}, the power consumption can be reduced, which leads to improved EE.
	
	\section{Adaptive RF chain activation}
	\label{sec_ARFA}
	
	\begin{figure}[t]
		\centering
		\includegraphics[scale=0.4]{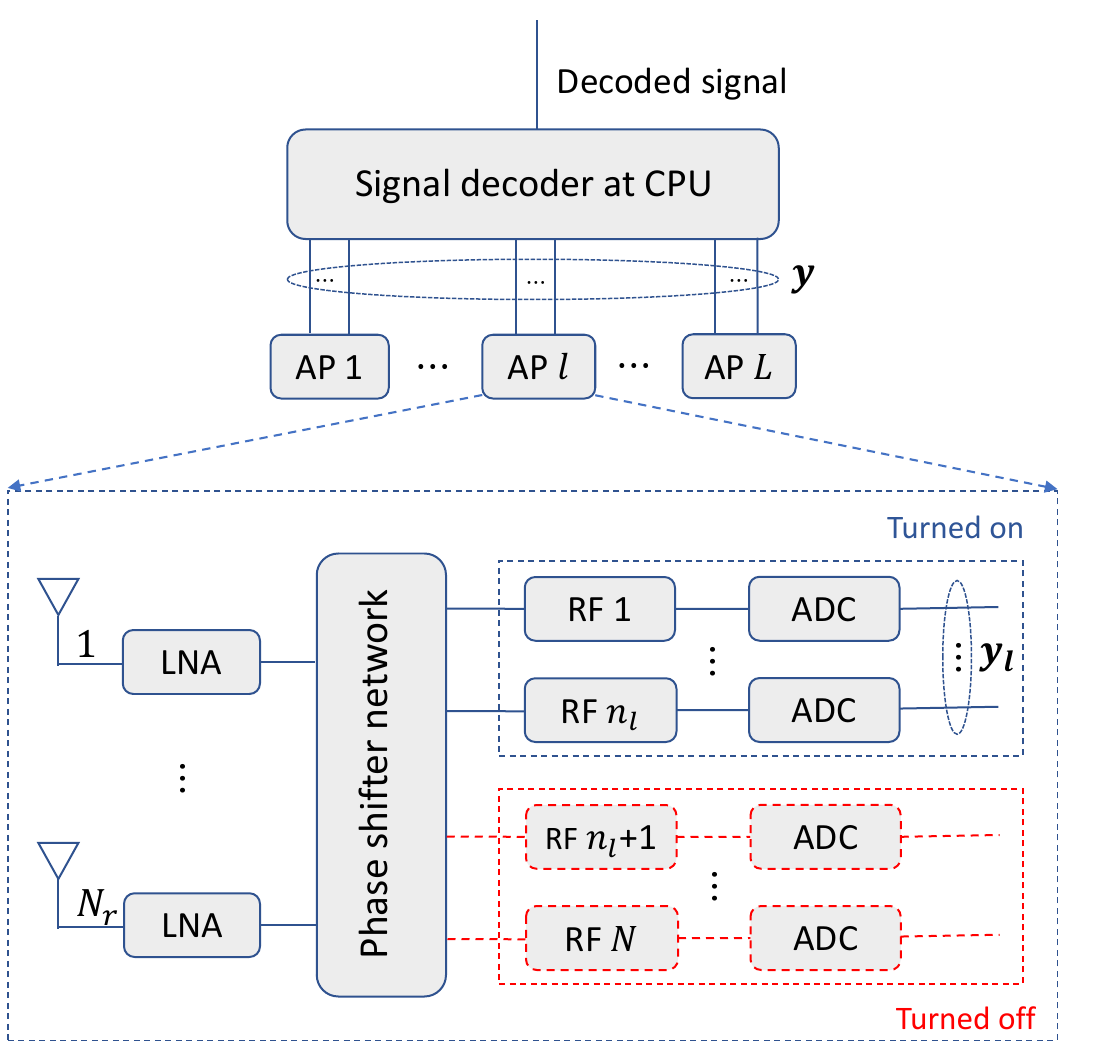}
		\caption{HBF architecture with the ARFA scheme. In the phase-shifter network, $n_l N_r$ out of $NN_r$ PSs are turned on at the $l$th AP.}
		\label{fig_ARFA_model}
	\end{figure}
	
	\subsection{Problem formulation and basic ideas}
	\label{sec_ARFA_problem}
	The global analog combiner $\mF$ can be expressed as
	\begin{align*}
	\mF = \text{diag} \{  \underbrace{[\vf_{11}, \ldots, \vf_{1N}]}_{\text {analog combiner at AP 1}}, \ldots, \underbrace{[\vf_{L1}, \ldots, \vf_{LN}]}_{\text {analog combiner at AP $L$}}  \},
	\end{align*}
	and the following facts are noted:
	\begin{itemize}
		\item Based on Theorem \ref{theorem_rate_C}, the total achievable rate {obtained by analog combining} can be expressed as a sum of $\{\bar{\mathcal{R}}^{\mathrm{a}}_1,\ldots,\bar{\mathcal{R}}^{\mathrm{a}}_L\}$ corresponding to APs $\{1,\ldots,L\}$. In \cfmimo systems, the APs are distributed in a large area, and their communication channels experience different path losses and shadowing effects. Therefore, the contributions of the local analog combiners at different APs to the total achievable rate {are of} different significances.
		
		\item In a local analog combiner $\mF_l$, combining vectors $\{\vf_{l1}, \ldots, \vf_{lN}\}$ have different contributions to the sub-rate $\bar{\mathcal{R}}^{\mathrm{a}}_l$ given in \eqref{R_n}. Specifically, the contribution of $\vf_{ln}$ is more significant than that of $\vf_{lm}$ if $n < m$ because $n$ and $m$ are the indices of the ordered singular values of {$\hat{\mH}_l \hat{\mQ}_{l-1}^{-1} \hat{\mH}_l^H$ in {SC-HBF} and of $\hat{\mH}_l$ in {D-HBF}}.
	\end{itemize}
	As a result, it is likely that a subset of analog combining vectors in $\{\vf_{11}, \ldots, \vf_{LN}\}$ are insignificant and can be removed from the global combiner $\mF$ without causing considerable performance loss. We note that at an AP, an analog combining vector represents the effect of $N_r$ PSs connected to an RF chain, followed by an ADC. Therefore, an insignificant analog combining vector can be removed from signal combining by turning off its corresponding RF chain, ADC, and PSs, which results in a reduction in the total power consumption. Motivated by this, we propose an ARFA scheme that selectively activates RF chains at the APs. Let $\vn = \{n_1, \ldots, n_L \}$, where $n_l$ is the number of turned-on RF chains out of $N$ RF chains installed at the $l$th AP, $0 \leq n_l \leq N$. We note that for $n_l=0$, all the RF chains at the $l$th AP are turned off, and this AP does not consume any power for signal combining. It is also noted that in the uplink training phase of the next coherent time, the deactivated RF chains are reactivated for channel estimation. The optimal activation of RF chains at the APs can be performed based on the following remark.
	
	\begin{remark}
		\label{rm_ARFA}
		Because $\vf_{ln}$ is always more important at the $l$th AP than $\vf_{lm}$ with $n < m$ in terms of achievable rate, the problem of optimal activation of RF chains at an AP is equivalent to finding an optimal number of turned-on RF chains at that AP. Specifically, for the $l$th AP, if the ARFA scheme suggests using $n_l^{\star}$ RF chains, then the first $n_l^{\star}$ RF chains corresponding to $\{\vf_{l1}, \ldots, \vf_{ln_l^{\star}}\}$ are selected for analog signal combining, whereas the others are deactivated to save power.
	\end{remark}
	
	The HBF architecture with the proposed ARFA scheme is illustrated in Fig.\ \ref{fig_ARFA_model}. As an example, at the $l$th AP, only $n_l$ out of $N$ RF chains are turned on. Furthermore, the ADCs and PSs connected to the inactive RF chains are also turned off. Consequently, the local combiner $\mF_l$ consists of only $n_l$ analog combining vectors, i.e., $\mF_l = \{ \vf_1, \ldots, \vf_{n_l} \}$.

	Unlike the conventional fixed-activation HBF, in the proposed ARFA scheme, the global combiner $\mF$, $\mathcal{R}^{\mathrm{a}}_l$, and {$\mathcal{R} \triangleq \sum_{l=1}^{L} \mathcal{R}^{\mathrm{a}}_l$} depend on $\vn$. Therefore, in this section, they are expressed as functions of $\vn$, i.e., $\mF(\vn)$, $\mathcal{R}^{\mathrm{a}}_l(\vn)$, and $\mathcal{R}(\vn)$, respectively. We limit the total number of turned-on RF chains in the ARFA scheme to $L\bar{n}$, i.e., $\sum_{l=1}^{L}n_l = L\bar{n}$, where $\bar{n} (\leq N)$ is the average number of activated RF chains at each AP. Based on Remark \ref{rm_ARFA}, the optimal activation of RF chains at the APs in the ARFA scheme can be performed by solving
	\begin{align*}
	\vnb = \arg \max_{\vn \in \mathcal{S}} \mathcal{R}(\vn), \numberthis \label{nbest}
	\end{align*}
	where $\mathcal{S} = \left\{ \vn: 0 \leq n_l \leq N, \sum_{l=1}^{L}n_l = L\bar{n} \right\}$ is the feasible set of $\vn$.
	The optimal $\vnb$ in \eqref{nbest} can be found by exhaustive search over the entire feasible set $\mathcal{S}$. However, in \cfmimo systems, $L$ is large; thus, an excessively large number of candidates for $\vn$ need to be examined in the exhaustive search, which is almost computationally prohibitive. In the following subsections, we propose three low-complexity algorithms to find $\vnb$. By abuse of notation, we use $\mF^{\star}$ for the global combiner found by the proposed ARFA algorithms. Furthermore, we note that Algorithm \ref{algorithm_C_HBF} can be easily modified for the ARFA scheme by replacing $N$ in steps 3, 7, and 10 with $n_l$, reflecting a dynamic number of analog combining vectors for the local combiner $\mF_l$ at the $l$th AP.

	\subsection{{ARFA with SC-HBF (SC-ARFA)}}
	
	\begin{algorithm}[t]
		\small
		\caption{{HBF with SC-ARFA}}
		\label{al_GSARFA}
		\begin{algorithmic}[1]
			\ENSURE $\mF^{\star}$
			\STATE Initialize $\vn = [n_1, \ldots, n_L]$ with $n_l = \bar{n}, \forall l$.
			\STATE Use Algorithm \ref{algorithm_C_HBF} to find $\mF(\vn)$, $\mathcal{R}^{\mathrm{a}}_l(\vn), \forall l$, and $\mathcal{R}(\vn)$.
			\STATE $\mF^{\star} = \mF(\vn)$, $\mathcal{R}^{\star} = \mathcal{R}(\vn)$.
			\STATE Obtain $\{n_{[1]}, \ldots, n_{[L]}\}$ s.t. $\mathcal{R}^{\mathrm{a}}_{[1]}(\vn) > \ldots > \mathcal{R}^{\mathrm{a}}_{[L]}(\vn)$.
			
			\STATE $i = 1$, $k=L$ 
			\WHILE {$i < k$}
			
			\WHILE {$n_{[i]} = N$}
			\STATE $i = i +1$
			\ENDWHILE
			\WHILE {$n_{[k]} = 0$}
			\STATE $k = k-1$
			\ENDWHILE

			\STATE Update $\vn$: $n_{[i]} = n_{[i]} + 1$, $n_{[k]} = n_{[k]} - 1$.
			\STATE Use Algorithm \ref{algorithm_C_HBF} to find $\mF(\vn)$ and $\mathcal{R}(\vn)$ with the updated $\vn$.
			\STATE Update $\mF^{\star} = \mF(\hat{\vn}^{\star})$ if $\mathcal{R}(\hat{\vn}^{\star}) > \mathcal{R}^{\star}$.

			\ENDWHILE
			\STATE {At the $l$th AP, compute $\mW_l^{\star}$ based on \eqref{eq_WMMSE}, $\forall l$}.
		\end{algorithmic}
	\end{algorithm}

	In {SC-ARFA}, the ARFA scheme is incorporated with {SC-HBF}, and the optimal numbers of active RF chains at the APs are found at the CPU based on the global CSI. The idea to find $\vn^{\star}$ is to turn on/off as many RF chains as possible at the APs corresponding to the largest/smallest $\mathcal{R}^{\mathrm{a}}_l$, as presented in Algorithm \ref{al_GSARFA}. In steps 1--3, all elements of $\vn$ are set to $\bar{n}$, then $\mF(\vn), \mathcal{R}^{\mathrm{a}}_l(\vn), \forall l$, and $\mathcal{R}(\vn)$ are computed. In step 4, the elements of $\vn$ are ordered to obtain $\{n_{[1]}, \ldots, n_{[L]}\}$ in the {decreasing} order of sub-rates $\{\mathcal{R}^{\mathrm{a}}_1(\vn), \ldots, \mathcal{R}^{\mathrm{a}}_L(\vn)\}$.  Therefore, $n_{[i]}$ is the number of turned-on RF chains at the AP with the $i$th largest sub-rates, i.e., $\mathcal{R}^{\mathrm{a}}_{[i]} (\vn)$. 
	
	In step 5, we initialize $i=1$ and $k=L$. In steps 6--16, $n_{[i]}$ is increased by one, whereas $n_{[k]}$ is decreased by one, in each iteration. We note that in step 13, $n_{[i]}$ and $n_{[k]}$ are updated simultaneously to guarantee $\sum_{l=1}^{L} n_{[l]} = L \bar{n}$. The updates of $n_{[i]}$ and $n_{[k]}$ result in a new candidate $\vn$. Hence, $\mF(\vn)$ and $\mathcal{R}(\vn)$ are found in step 14, and $\mF^{\star}$ is updated if the performance is improved, as shown in step 15. Once $n_{[i]}$ reaches the maximum, i.e., $N$, the number of turned-on RF chains at the AP associated with the $(i+1)$th largest sub-rate, i.e., $n_{[i+1]}$, is considered, as shown in steps 7--9. In contrast, if $n_{[k]}$ reaches the minimum, i.e., zero, $n_{[k-1]}$ is considered next, as shown in steps 10--12. This iterative process is terminated if $i \geq k$, for which we have $\mathcal{R}^{\mathrm{a}}_{[i]} (\vn) \leq \mathcal{R}^{\mathrm{a}}_{[k]} (\vn)$ and the increase (decrease) in $n_{[i]}$ $(n_{[k]})$ is unlikely to provide performance improvement. {Once all the analog combiners are determined and sent to the APs, the digital combiner at each AP is determined, as in step 17.}
	
	{We note that the ARFA process needs to be performed at the CPU to jointly optimize the numbers of RF chains at all APs. In the {SC-ARFA} schemes, the global CSI is exploited to evaluate the candidates for $\vn$. However, the employment of {SC-HBF} in these schemes requires high computational complexity and a large amount of information exchanged between the CPU and APs, as discussed in Section III-B. This motivates us to propose an ARFA scheme incorporated with {D-HBF} in the next subsection.}
	
	\subsection{{ARFA with D-HBF (D-ARFA)}}
	
	Without global CSI, the ARFA scheme can be performed if the CPU knows the qualities of the available combining vectors or the path loss corresponding to each AP. The former idea relies on the fact that a combining vector is obtained by quantizing a singular vector of the channel matrix, as shown in \eqref{f_tilde}. Therefore, the quality of a combining vector can be evaluated based on its corresponding singular value. In contrast, the latter idea for {D-ARFA} is motivated by the observation that the AP with more significant path loss should have fewer activated RF chains because it is more likely to have a low sub-rate.
	
	\subsubsection{{Singular values-based D-ARFA (SV-based D-ARFA)}}
	
	{The SV-based {D-ARFA} scheme is summarized in Algorithm \ref{al_SV_SCARFA}. Specifically, in step 1, each AP finds and sends the $N$ largest singular values of the channel matrix to the CPU. Here, only the $N$ largest singular values are sent because, in the proposed ARFA scheme, only $n_l$ out of $N$ combining vectors are selected for signal combining at the $l$th AP. As a result, the set of $LN$ singular values $\left\{\lambda_1^{(1)}, \ldots, \lambda_N^{(1)}, \ldots, \lambda_1^{(L)}, \ldots, \lambda_N^{(L)}\right\}$ is available at the CPU, where $\lambda_n^{(l)}$ is the $n$th largest singular value associated with the $l$th AP. Then, the numbers of active RF chains at the APs are determined in steps 3--6. Specifically, an RF chain at an AP is suggested for activation if its corresponding singular value is not smaller than $\lambda_{L\bar{n}}$ found in step 2. In other words, the number of active RF chains at the $l$th AP, that is, $n^{\star}_l$, is set as the number of elements in $\left\{\lambda_1^{(l)}, \ldots, \lambda_N^{(l)}\right\}$ that are not smaller than $\lambda_{L\bar{n}}$. Finally, the CPU sends the value $n^{\star}_l$ back to the $l$th AP, which is then used for signal combining based on the {D-HBF} scheme.}
	
	\begin{algorithm}[t]
		\small
		\caption{{HBF with SV-based D-ARFA}}
		\label{al_SV_SCARFA}
		\begin{algorithmic}[1]
			\ENSURE $\mF^{\star}$
			\STATE Each AP finds the $N$ largest singular values of its channel matrix and sends them to the CPU. Specifically, the $l$th AP finds and sends a vector $\ve_l = \left[\lambda_1^{(l)}, \ldots, \lambda_N^{(l)}\right]$, where $\lambda_n^{(l)}$ is the $n$th largest singular value of $\hat{\mH}_l$.
			\STATE The CPU finds $\lambda_{L\bar{n}}$, which is the $(L \bar{n})$th largest element in the singular value set $\{\lambda_1^{(1)}, \ldots, \lambda_N^{(1)}, \ldots, \lambda_1^{(L)}, \ldots, \lambda_N^{(L)} \}$ received from all APs.
			
			\FOR{$l = 1 \rightarrow L$}
			\STATE The CPU sets $n^{\star}_l$ to the number of elements in $\ve_l$ that are not smaller than $\lambda_{L\bar{n}}$ and sends $n^{\star}_l$ to the $l$th AP.
			\STATE The $l$th AP determines its local analog combiner $\mF_l^{\star}$ for $n^{\star}_l$ RF chains, i.e., $\mF_l^{\star} = [\tilde{\vf}_{l1}^{\star}, \ldots, \tilde{\vf}_{ln^{\star}_l}^{\star}]$, where $\tilde{\vf}_{ln}^{\star}$ is given by \eqref{f_tilde}, {and determines $\mW_l^{\star}$ based on \eqref{eq_WMMSE}.}
			\ENDFOR
		\end{algorithmic}
	\end{algorithm}
	
	\subsubsection{{Path loss-based {D-ARFA} (PL-based D-ARFA})}
	
	\begin{algorithm}[t]
		\small
		\caption{{HBF with PL-based {D-ARFA}}}
		\label{al_PL_SCARFA}
		\begin{algorithmic}[1]
			\ENSURE {$\mF^{\star}$}
			\STATE {Find $\vn = \{n_1, \ldots, n_L \}$ based on \eqref{vn_raw}.}
			\STATE {Obtain $\{n_{[1]}, \ldots, n_{[L]}\}$ s.t. $\alpha_{[1]} > \ldots > \alpha_{[L]}$.}
			\STATE {$t = 1$}
			\WHILE {{$\vn \notin \mathcal{S}$}}
			\IF {{$\sum_{l=1}^{L} n_{[l]} < L\bar{n}$ and $n_{[t]} < N$}}
			\STATE {$n_{[t]} = n_{[t]} + 1$}
			\ENDIF 
			\IF {{$\sum_{l=1}^{L}n_{[l]} > L\bar{n}$ and $n_{[L-t+1]} > 0$}}
			\STATE {$n_{[L-t+1]} = n_{[L-t+1]} - 1$}
			\ENDIF
			
			\STATE {$t=t+1$}
			\STATE {Reset $t = 1$ if $t > L$.}
			\ENDWHILE
			
			\STATE {Obtain $\vnb$ by reordering $\{n_{[1]}, \ldots, n_{[L]}\}$ to the original order.}
			\FOR {{$l = 1 \rightarrow L$}}
			\STATE {The CPU sends $n_l^{\star}$ to the $l$th AP.}
			\STATE The $l$th AP determines its local analog combiner $\mF_l^{\star}$ for $n^{\star}_l$ RF chains, i.e., $\mF_l^{\star} = [\tilde{\vf}_{l1}^{\star}, \ldots, \tilde{\vf}_{ln^{\star}_l}^{\star}]$, where $\tilde{\vf}_{ln}^{\star}$ is given by \eqref{f_tilde} {and determine $\mW_l^{\star}$ based on \eqref{eq_WMMSE}}.
			\ENDFOR
			
		\end{algorithmic}
	\end{algorithm}
	
	In the SV-based D-ARFA scheme, the largest singular values of the channel matrices are required to find $\vnb$. This entails high computational complexity, especially when $L$ and $N_r$ are large. To avoid this, we herein propose the PL-based D-ARFA scheme, in which $\vnb$ is obtained based on the total path losses associated with the APs. In CF massive MIMO systems, the APs are distributed in a large area. Therefore, the contribution of an AP to the total achievable rate considerably depends on its path loss.
	
	Let $\beta_l = \sum_{k=1}^{K} \beta_{kl}$ be the sum of path loss of the $l$th AP, with $\beta_{kl}$ given in \eqref{pathloss}, and let $\alpha_l = \frac{1}{\beta_l}, \forall l$. The number of activated RF chains at the $l$th AP can be set to
	\begin{align*}
	n_l = \min \left\{N, \round*{L\bar{n} \frac{\alpha_l}{\sum_{i=1}^{L} \alpha_i}} \right\}, \forall l, \numberthis \label{vn_raw}
	\end{align*}
	where $\min \{N, \cdot\}$ is used to guarantee $n_l \leq N$, and $\round{\cdot}$ rounds a real number to its nearest integer. However, because of rounding, it is possible to obtain $\sum_{l=1}^{L}n_l \neq L\bar{n}$, which leads to $\vn \notin \mathcal{S}$. To solve this problem, we propose Algorithm \ref{al_PL_SCARFA}.
	
	In Algorithm \ref{al_PL_SCARFA}, the elements of $\vn$ found in step 1 based on \eqref{vn_raw} are sorted in step 2 in decreasing order of $\{ \alpha_1, \ldots, \alpha_L \}$, i.e., in increasing order of the sums of path loss $\{ \beta_1, \ldots, \beta_L \}$, to generate $\{n_{[1]}, \ldots, n_{[L]}\}$. Here, the order index $[t]$ indicates that $n_{[t]}$ RF chains are chosen to be activated at the AP associated with the $t$th-smallest path loss. Therefore, if $\sum_{l=1}^{L} n_{[l]} < L\bar{n}$ and $n_{[t]} < N$, $n_{[t]}$ is increased by one. In contrast, if $\sum_{l=1}^{L}n_{[l]} > L\bar{n}$ and $n_{[L-t+1]} > 0$, $n_{[L-t+1]}$ is decreased by one. This process is repeated until $\vn \in \mathcal{S}$ is satisfied, as shown in steps 3--13. In this procedure, by initializing $t=1$ and gradually increasing $t$, the numbers of turned-on RF chains for the APs with path loss are chosen to increase first, whereas those for the APs with larger path loss are chosen to decrease first. In step 14, $\{n_{[1]}, \ldots, n_{[L]}\}$ are reordered into the original order. In steps 15--18, the numbers of active RF chains at the APs are determined, which are then fed back to the APs for SC-HBF, as in steps 3--6 of Algorithm \ref{al_SV_SCARFA}.
	
	\section{{Power consumption analysis}}
	\label{sec_power}
	
	{In the considered uplink CF mMIMO system, the total power consumption is modeled as \cite{dai2016energy, bashar2019energy, bjornson2015optimal, payami2018phase}
		\begin{align*}
		P_{\text{total}} = \sum_{k=1}^{K} \left(P_{\text{TX},k} + P_{\text{UE},k}\right) + \sum_{l=1}^{L} \left(P_{\text{fix},l} + P_{\text{BF},l} + P_{\text{FH},l}\right), \numberthis \label{Ptot}
		\end{align*}
		where $P_{\text{TX},k}$ and $P_{\text{UE},k}$ represent the transmit power and the required power to run circuit components at the $k$th UE, respectively; $P_{\text{fix},l}$, $P_{\text{BF},l}$, and $P_{\text{FH},l}$ respectively denote the fixed power consumption term, the variable power consumption for the beamforming structure, and the fronthaul power consumption for the $l$th AP. $P_{\text{TX},k}$ is given as 
		\begin{align*}
		P_{\text{TX},k} = \gamma \sigma^2 \sum_{k=1}^{K} \frac{1}{\eta_k} \mean{\norm{\vx_k}^2} =  \sum_{k=1}^{K} \frac{\gamma \sigma^2}{\eta_k}, \numberthis \label{Ptx}
		\end{align*}
		where $\eta_k \in (0,1]$ denotes the power amplifier efficiency of the UE $k$, and the last equality is obtained by $\mean{\norm{\vx_k}^2} = 1, \forall k$. In an HBF architecture, each antenna requires a low-noise amplifier (LNA) and two mixers, and each RF chain requires one ADC and $N_r$ PSs, as illustrated in Fig.\ \ref{fig_ARFA_model}  \cite{nguyen2019unequally, roth2017achievable, roth2018comparison}. Therefore, $P_{\text{BF},l}$ linearly depends on the numbers of antennas $(N_r)$ and active RF chains at the $l$th AP $(n_l)$ as follows:
		\begin{align*}
		P_{\text{BF},l} = N_r p_{\text{BF},1} + n_l p_{\text{BF},2}, \numberthis \label{Pbf}
		\end{align*}
		where $p_{\text{BF},1} = p_{\text{LNA}} + 2p_{\text{M}}$, $p_{\text{BF},2} =N_r p_{\text{PS}} + p_{\text{RF}} + p_{\text{ADC}}$, with $p_{\text{LNA}}$, $p_{\text{M}}$, $p_{\text{PS}}$, $p_{\text{RF}}$, and $p_{\text{ADC}}$ respectively denoting the power consumed by an LNA, mixer, PS, RF chain, and ADC. Furthermore, $P_{\text{FH},l}$ can be obtained by \cite{dai2016energy, bashar2019energy}
		\begin{align*}
		P_{\text{FH},l} = P_{\text{FH,max}} \frac{R_{\text{FH},l}}{C_{\text{FH},l}} = \kappa_l R_{\text{FH},l}, \numberthis \label{Pfh} 
		\end{align*}}
	{where $P_{\text{FH,max}}$ is the maximum power required for the fronthaul traffic at the full capacity $C_{\text{FH},l}$, $\mathcal{R}_{\text{FH},l}$ is the actual fronthaul rate between the $l$th AP and the CPU, and $\kappa_l = \frac{P_{\text{FH,max}}}{C_{\text{FH},l}}$. In the considered decentralized signal processing scheme, $ 2 K  \tau_d \alpha_l$ bits are required to quantize the signal vector $\vr_l \in \setC^{K \times 1}$ during each coherence interval \cite{bashar2019energy, bashar2020uplink} at the $l$th AP before being sent to the CPU. Here, $\alpha_l$ is the number of quantization bits at the $l$th AP, and $\tau_d$ is the length (in symbols) of the uplink data. As a result, $\mathcal{R}_{\text{FH},l}$ is given by \cite{bashar2019energy, bashar2020uplink}
		\begin{align*}
		R_{\text{FH},l} = \frac{2K  \tau_d \alpha_l}{T_c}, \numberthis \label{Rfh}
		\end{align*}
		where $T_c$ is the coherence time (in seconds). Assume that all the UEs have the same power amplifier efficiency and circuit power consumption, i.e., $\eta_k = \eta$, $P_{\text{UE},k} = P_{\text{UE}}, \forall k$, and that all APs have the same fixed power consumption, number of quantization bits, and capacity, i.e., $P_{\text{fix},l} = P_{\text{fix}}$, $\alpha_l = \alpha$, $C_{\text{FH},l} = C_{\text{FH}}$, $\kappa_l = \kappa$, $\forall l$. Then, we have $P_{\text{FH},l} = P_{\text{FH}}$ and $\mathcal{R}_{\text{FH},l} = R_{\text{FH}}$, $\forall l$. Furthermore, we note that AP $l$ requires $P_{\text{fix}}$, even when it is in sleep mode; in contrast, $P_{\text{FH}}$ and $P_{\text{BF},l}$ are only consumed when it is in the active mode. Let $\setA$ be the set of APs in active mode and $\abs{\setA}$ be the number of active APs. Then, from \eqref{Ptot}--\eqref{Rfh}, the total power consumption can be expressed as
		\begin{align*}
		P_{\text{total}} &= \frac{K \gamma \sigma^2}{\eta} + K P_{\text{UE}} + L P_{\text{fix}} +  \abs{\setA}  P_{\text{FH}}  + \sum_{l \in \setA} \left(N_r p_{\text{BF},1} + n_l p_{\text{BF},2}\right), \\
		&= P_0 +  \abs{\setA}  P_{\text{FH}} + \abs{\setA} N_r p_{\text{BF},1} + p_{\text{BF},2} \sum_{l \in \setA} n_l, \numberthis \label{Ptot_1}
		\end{align*}
		where $P_0 = \frac{K \gamma \sigma^2}{\eta} + K P_{\text{UE}} + L P_{\text{fix}}$, a fixed term in $P_{\text{total}}$, for simple exposition.}
	
	{It is observed from \eqref{Ptot_1}  that $P_{\text{total}}$ varies depending on the number of active APs, i.e., $\abs{\setA}$; the total number of turned on RF chains, i.e., $\sum_{l \in \setA} n_l$; and the number of antennas $N_r$. More specifically, it is a linearly increasing function of these factors. Therefore, $P_{\text{total}}$ can be minimized by using only a subset of APs in the APS scheme \cite{ngo2017total}, using a reduced number of antennas in the AS scheme \cite{tai2015low}, or optimizing both  $\sum_{l \in \setA} n_l$ and $\abs{\setA}$ in the proposed ARFA scheme. Next, we compare these schemes in terms of the total power consumption. Furthermore, conventional fixed-activation HBF schemes are also considered as benchmarks.}
	
	{$\bullet$ \emph{ARFA scheme:}
		When the ARFA scheme is employed, $n_l$ is different among the APs; however, the total number of RF chains is fixed to $L\bar{n}$, i.e., $\sum_{l \in \setA} n_l = L\bar{n}$. By inserting this into \eqref{Ptot_1}, we obtain
		\begin{align*}
		P_{\text{total}}^{\text {ARFA}} = P_0 + \abs{\setA}  P_{\text{FH}}   + \abs{\setA} N_r p_{\text{BF},1} + L\bar{n} p_{\text{BF},2}, \numberthis \label{P_ARFA}
		\end{align*}
		where $\setA$ contains only the APs with at least one activated RF chain. Therefore, we have $\abs{\setA} = \sum_{l \in \setA} \delta_{l}$  with $\delta_{l} = 1$ if $n_l > 0$, and $\delta_{l} = 0$ if $n_l = 0$. We note that the proposed ARFA algorithms have different operations, which can result in different $\setA$. Therefore, they can have different power consumption.}
	
	{$\bullet$ \emph{Fixed-activation HBF:}
		We refer to the {SC-HBF and D-HBF} without the ARFA as the \textit{fixed-activation HBF}. In this scheme, the same number of RF chains are activated at all $L$ APs. For comparison with the proposed ARFA schemes, we consider two deployments: $n_l=N, \forall l$ and $n_l=\bar{n}, \forall l$. We note that with fixed activation HBF, all the APs are in active mode because they have a fixed nonzero number of RF chains for signal processing, i.e., $\abs{\setA} = L$. By inserting $n_l=N, \forall l$, and $n_l=\bar{n}, \forall l$ into \eqref{Ptot_1}, we obtain
		\begin{align*}
		P_{\text{total}}^{\text{fix},N} = P_0 +  L P_{\text{FH}} +  L N_r p_{\text{BF},1} + L N p_{\text{BF},2}, \numberthis \label{P_conv_N}\\
		P_{\text{total}}^{\text{fix},\bar{n}} = P_0 +  L P_{\text{FH}} +  L N_r p_{\text{BF},1} + L \bar{n} p_{\text{BF},2} . \numberthis \label{P_conv_nbar}
		\end{align*}}
	{$\bullet$ \emph{APS scheme:}
		{In this scheme, only a subset of the APs is selected based on received power \cite{ngo2017total}, whereas the others are put into the sleep mode}. For comparison with the proposed schemes in mmWave systems, we assume the conventional deployment of RF chains at the APs, i.e., each AP is equipped with $N$ RF chains, all of which are used for analog signal combining, i.e., $n_l = N, \forall l$. For a fair comparison, the number of APs in active mode in this scheme is assumed to be $\frac{L\bar{n}}{N}$. This guarantees that a total of $L\bar{n}$ RF chains are used at the selected APs, which is equal to the	number of activated RF chains in the proposed ARFA scheme and fixed-activation HBF scheme with $n_l = \bar{n}, \forall l$. By inserting $n_l = N, \forall l$, and $\abs{\setA} = \frac{L\bar{n}}{N}$ into \eqref{Ptot_1}, we have
		\begin{align*}
		P_{\text{total}}^{\text{APS}} = P_0 +  \frac{L\bar{n}}{N} P_{\text{FH}} +  \frac{L\bar{n}}{N} N_r p_{\text{BF},1} + L\bar{n} p_{\text{BF},2}. \numberthis \label{P_APS}
		\end{align*}}
	{$\bullet$ \emph{AS scheme:}
		In the AS scheme, at each AP, only $N_r^{\text{AS}}$ of $N_r$ antennas are activated, corresponding to $N_r^{\text{AS}}$ received signals put through the digital signal combining \cite{tai2015low}. In other words, analog signal combining is conducted by a network of $N_r^{\text{AS}}$ switches rather than $N N_r$ PSs, in contrast to the other compared schemes. Therefore, at each AP, $N_r$ switches are required, whereas the numbers of antennas, RF chains, and ADCs are the same and as small as $N_r^{\text{AS}}$, and the number of mixers is $2N_r^{\text{AS}}$. Furthermore, in the AS scheme, all the APs are in the active mode, i.e., $\abs{\setA} = L$. Let $p_{\text{SW}}$ be the power consumed by a switch. The total power consumption in this scheme is given as
		\begin{align*}
		P_{\text{total}}^{\text {AS}} = P_0 + L  P_{\text{FH}} + L N_r p_{\text{SW}} + L N_r^{\text{AS}} (p_{\text{RF}} + p_{\text{ADC}} + p_{\text{BF},1}). \numberthis \label{P_AS}
		\end{align*}}

	By comparing \eqref{P_ARFA} to \eqref{P_conv_N}--\eqref{P_APS}, we observe that:
	\begin{itemize}
		\item The proposed ARFA scheme requires no higher power consumption than the fixed-activation HBF schemes with $n_l = N$ and $n_l = \bar{n}, \forall l$, because $\bar{n} < N$ and $\abs{\setA} \leq L$. {Furthermore, we note that a dominant part of the power consumed for beamforming is created by the RF chains and ADC. Therefore, from \eqref{P_ARFA} and \eqref{P_conv_N}, it is clear that a considerable reduction in power consumption can be obtained by the ARFA if $\bar{n} \ll N$ is chosen.}
		
		\item Both the power consumption and total achievable rate of the proposed ARFA scheme significantly depend on $\bar{n}$. Specifically, a smaller $\bar{n}$ leads to a reduction in both power consumption and total achievable rate with respect to the fixed-activation HBF scheme with $n_l = N, \forall l$. This tradeoff is discussed further in the next section.
		
		\item It is observed from \eqref{P_APS} and \eqref{P_ARFA} that the APS and proposed ARFA schemes have a difference of {$\abs{\frac{L\bar{n}}{N} - \abs{\setA}} (P_{\text{FH}} + N_r p_{\text{BF,1}} )$} in power consumption, even though they have the same total number of active RF chains. Specifically, the APS scheme requires slightly lower power consumption, but its achievable rate is much lower than that of the ARFA scheme, as is shown in the next section. {It is not certain from \eqref{P_AS} and \eqref{P_ARFA} which of AS and ARFA schemes has the lower power consumption, which will be determined based on the simulation results in the next section.}
	\end{itemize}
	
	\section{Simulation results}
	
	\label{sec_sim_result}

	\subsection{{Simulation parameters}}
	
	\begin{table*}[t]
		\vspace{-1cm}
		\renewcommand{\arraystretch}{0.8}
		\caption{Simulation parameters \cite{bashar2019energy,ngo2017total,nguyen2019unequally,dai2016energy}}
		\label{tab_sim_params}
		\centering
		\begin{tabular}{|c|c|}
			\hline
			
			Parameters    &  Values\\
			\hline
			\hline
			Power amplifier efficiency & $\eta = 0.3$ \\
			\hline
			Coherent time and data length & $T_c = 2$ ms, $\tau_c = 200$, $\tau_p = 20$ symbols \\
			\hline
			No. of quantization bits & $\alpha = 2$ bits \\
			\hline
			UE and fixed power term & \makecell{$P_{\text{UE}} = 1$ W, $P_{\text{fix}} = 0.825$ W } \\
			\hline
			Fronthaul capacity and required power & \makecell{$C_{\text{FH}} = 100$ Mbps, $P_{\text{FH,max}} = 50$ W} \\
			\hline
			Component power & \makecell{$p_{\text{LNA}} = 20$ mW, $p_{\text{ADC}} = 200$ mW, $p_{\text{RF}} = 40$ mW \\ $p_{\text{PS}} = 30$ mW, $p_{\text{M}} = 0.3$ mW, $p_{\text{SW}} = 5$ mW, $\rho_p = 100$ mW} \\
			\hline
		\end{tabular}
	\end{table*}
	Simulations are performed to evaluate the total achievable rates, power consumption, EEs, and computational complexities of the proposed {SC-HBF, D-HBF}, and ARFA schemes. In simulations, $K$ UEs and $L$ APs are uniformly distributed at random within a square coverage area of size $D \times D$ $\text{m}^2$, where $D$ is set to $1000$ m \cite{ngo2017cell}. The large-scale fading coefficients are computed based on \eqref{pathloss} with $\epsilon = 4.1$, $\xi = 7.6$, and the antenna gain is set to $G_{\mathrm{a}} = 15$ dBi \cite{rappaport2015wideband}.  Furthermore, we assume $f_c = 28$ GHz, $B = 100$ MHz, and $\text{NF} = 9$ dB for the carrier frequency, system bandwidth, and noise figure, respectively. As a result, the noise power is given as $\sigma^2 = -174 \text{ dBm/Hz} + 10 \log_{10} (B) + \text{NF}$.
	
	The channel coefficients between each UE and AP are generated based on the geometric Saleh--Valenzuela channel model given in \eqref{eq_h}. For simplicity, we assume an identical number of effective channel paths between each UE and AP, which is set to $P_{kl} = 20, \forall l,k$ \cite{alkhateeb2015limited, alkhateeb2014channel, nguyen2018coverage}, reflecting the limited scattering in mmWave channels. {The AoAs are uniformly distributed in $\left[ -\frac{\pi}{12}, \frac{\pi}{12} \right]$. The ULA model is employed for the antenna arrays at the APs and UEs with antenna spacing of half a wavelength, i.e., $\frac{d_s}{\lambda} = \frac{1}{2}$ \cite{gao2016energy, bogale2016number}. The phases in the analog combiner are selected from $ \Theta = \left\{ 0, \frac{2\pi}{2^b}, \frac{4\pi}{2^b}, \ldots, \frac{2(2^b-1)\pi}{2^b} \right\}$, where $b=4$ is set, implying 4-bit quantization of the PSs. The parameters in Table \ref{tab_sim_params} are assumed to compute the total power consumption \cite{bashar2019energy,ngo2017total,nguyen2019unequally,dai2016energy}.

	\subsection{Performance of the C-HBF and SC-HBF schemes}
	
	We numerically evaluate the total achievable rates of the C-HBF and SC-HBF schemes, which are analyzed in Section \ref{sec_HBF}. We assume the conventional RF chain deployment in this section, i.e., all $N$ available RF chains are active for analog combining. For comparison, we consider the beam selection scheme, in which the analog beamforming vectors are selected from a discrete Fourier transform (DFT) codebook based on an exhaustive search \cite{han2018dft,yang2010dft}. Fig.\ \ref{fig_rate_HBF} shows the total achievable rates of the SC-HBF, D-HBF, and beam selection schemes with $(N_r, K)=\{ (32,4), (64,8) \}$, $N=K$, and $L=32$ \cite{li2018performance}. It is clear from Fig.\ \ref{fig_rate_HBF} that for both the considered systems, the SC-HBF and D-HBF schemes have almost the same total achievable rates because they employ the same digital beamformer and their analog beamformers perform approximately the same, as stated in Remark \ref{remark_rate_SC}. Specifically, the former performs only slightly better than the latter. It is also observed in Fig.\ \ref{fig_rate_HBF} that the proposed SC-HBF and D-HBF schemes outperform the beam selection method for both the considered systems. 
	
	\begin{figure}[t]
		\centering
		\includegraphics[scale=0.52]{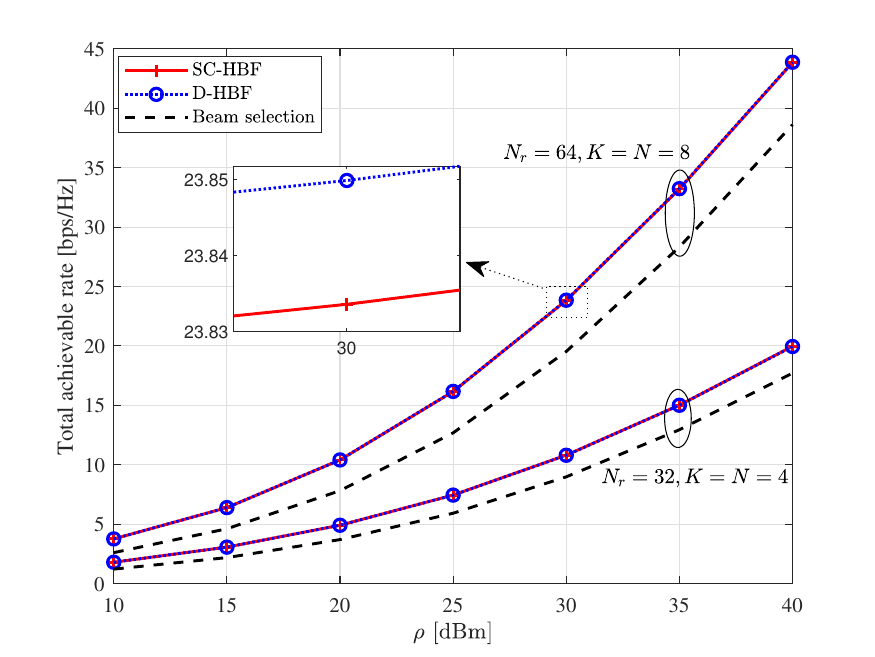}
		\caption{Total achievable rates of the C-HBF and SC-HBF schemes compared to those of the beam selection scheme with $L=32$, $K=8$, $N_r = 64$, and $N = 8$.}
		\label{fig_rate_HBF}
	\end{figure}

	\subsection{Performance of the proposed ARFA scheme}

	The total achievable rates, power consumption, and EEs of the proposed ARFA schemes, namely, {SC-ARFA}, {PL-based {D-ARFA}, and SV-based {D-ARFA},} are compared to those of the fixed-activation HBF with $n_l=N$ and $n_l=\bar{n}, \forall l$, APS, and AS schemes discussed in Section \ref{sec_power}. In our simulations, {SC-HBF} is used for the fixed-activation HBF and APS schemes. We note that the {SC- and D-HBF} provides almost identical performance, as shown in Fig.\ \ref{fig_rate_HBF}, and for the same RF chain deployment, they have the same power consumption. We consider a \cfmimo system with $L=32$, $K=8$, $N_r=64$, $N=8$ \cite{femenias2019cell, li2018performance}, and $\bar{n}=2$. In the AS scheme, the number of selected antennas at each AP is set to $N_r^{\text{AS}} = 32$, which ensures that the AS scheme achieves comparable total achievable rates with respect to the proposed schemes, allowing us to compare them in terms of EE. In the simulations, the power consumption of the fixed-activation HBF schemes with $n_l=N$, $n_l=\bar{n}$, the APS, and AS scheme is computed based on \eqref{P_conv_N}--\eqref{P_AS}, whereas that of the proposed ARFA schemes is obtained through simulations because it depends on $\delta_{n_l}$, as indicated in \eqref{P_ARFA}. The EE of a scheme is calculated as the ratio between the total achievable rate and the total power consumption. Furthermore, for a fair comparison, we fix the total number of active RF chains for each compared scheme to $L \bar{n}$, which ensures that an average of $\bar{n}$ RF chains are activated at each AP in all compared schemes.
	
	\begin{figure}[t]
		\subfigure[Total achievable rate]
		{
			\includegraphics[scale=0.5]{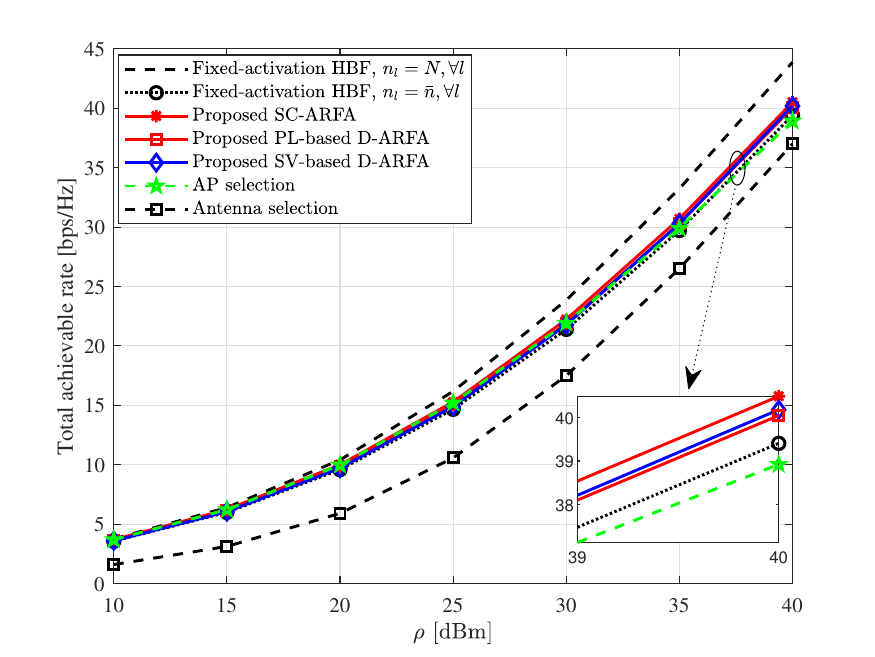}
			\label{fig_rate_40}
		}
		\subfigure[Energy efficiency]
		{
			\includegraphics[scale=0.5]{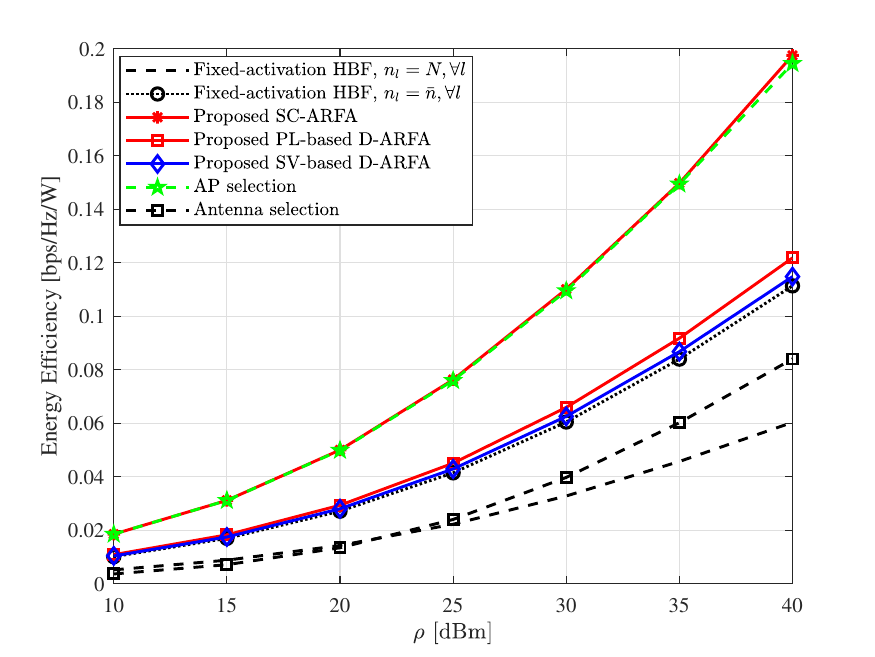}
			\label{fig_EE_40}
		}
		\caption{Total achievable rates and EEs of the proposed ARFA schemes compared to those of the fixed-activation HBF with $n_l=N$, $n_l=\bar{n}, \forall l$, APS, {and AS} schemes. Simulation parameters are $L=32$, $K = 8$, $N_r = 64$, $N = 8$, and $\bar{n}=2$.}
		\label{fig_perf_vs_Pt}
	\end{figure}
	
	In Fig.\ \ref{fig_perf_vs_Pt}, we show the total achievable rates and EEs of the considered schemes versus the average transmit power $\rho$ for $L=32$, $K = 8$, $N_r = 64$, $N = 8$, and $\bar{n}=2$. From Fig.\ \ref{fig_perf_vs_Pt}, the following observations are noted:
	\begin{itemize}
		\item It is seen that the fixed-activation HBF scheme with $n_l=N$ achieves the highest total achievable rate, as seen in Fig.\ \ref{fig_rate_40}, because it activates all the available APs and RF chains. However, in this scheme, high power is consumed by $LN$ RF chains. Therefore, its EE is significantly lower than those of the other considered schemes, in which only $L\bar{n}$ $(\ll LN)$ RF chains are turned on, as seen in Fig.\ \ref{fig_EE_40}. 
		\item Among the proposed ARFA schemes, the SC-ARFA achieves the highest achievable rate and EE. However, all of these schemes achieve remarkable improvement in EE with a small loss in the total achievable rate with respect to the fixed-activation HBF scheme with $n_l=N$..
		\item It is also observed that the proposed ARFA schemes outperform the AS scheme in terms of both the total achievable rate and EE. As it attains a comparable EE to the SC-ARFA scheme, the APS scheme is seen to be energy-efficient in Fig. \ref{fig_EE_40}, but its achievable rate is lower than those of all the proposed ARFA schemes. The fixed-activation HBF scheme with $n_l = \bar{n}$ is outperformed by the proposed SC-ARFA schemes in both achievable rate and EE.
	\end{itemize}

	\begin{figure*}[t]
		\subfigure[Total achievable rate]
		{
			\includegraphics[scale=0.38]{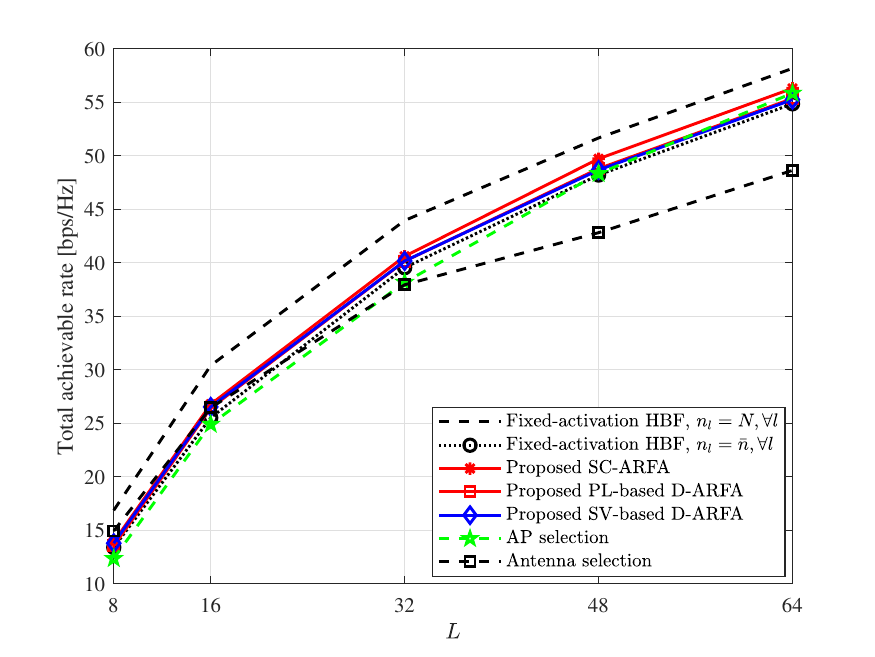}\hspace*{-2em}
			\label{fig_rate_vs_L}
		}
		\subfigure[Energy efficiency]
		{
			\includegraphics[scale=0.38]{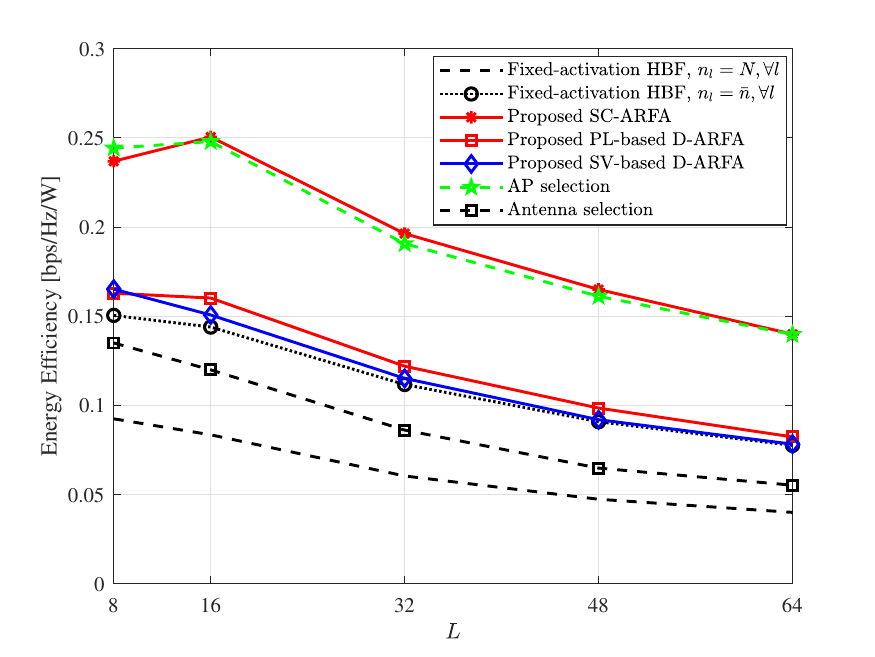}\hspace*{-2em}
			\label{fig_EE_vs_L}
		}
		\subfigure[Total power consumption]
		{
			\includegraphics[scale=0.38]{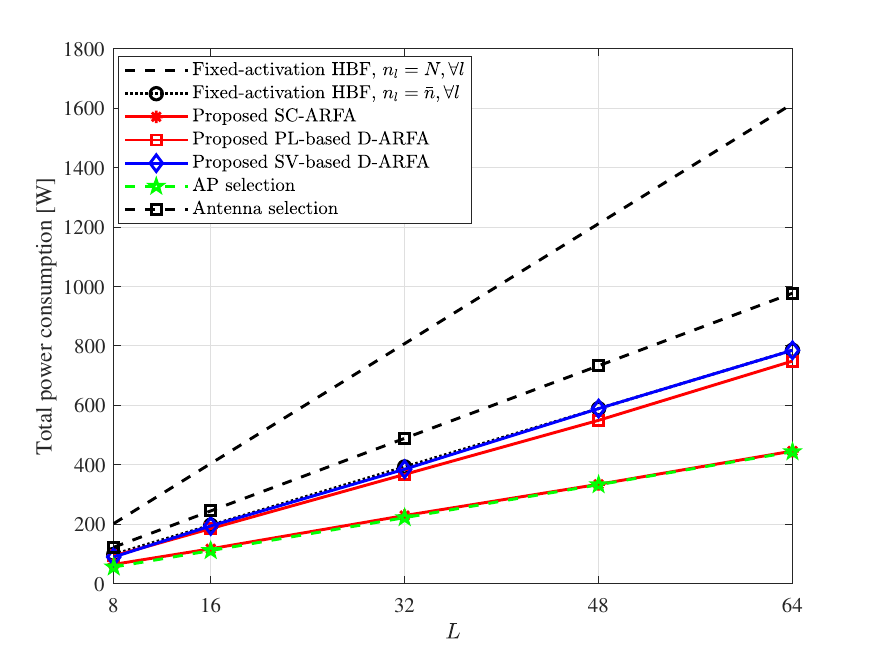}
			\label{fig_power_vs_L}
		}
		\caption{Total achievable rates, EEs, and {power consumption} of the proposed ARFA schemes compared to those of the fixed-activation HBF with $n_l=N$, $n_l=\bar{n}, \forall l$, APS, {and AS} schemes. Simulation parameters are $L=\{8,16,32,48,64\}$, $K = 8$, $N_r = 64$, $N = 8$, $\bar{n}=2$, and $\rho = 40$ dBm.}
		\label{fig_perf_vs_L}
	\end{figure*}

	In Fig.\ \ref{fig_perf_vs_L}, we show the total achievable rates, EEs, and {power consumption} of the considered schemes versus the number of APs. In this figure, we use the same simulation parameters as in Fig.\ \ref{fig_perf_vs_Pt}, except for the varying numbers of APs, i.e., $L=\{8,16,32,48,64\}$, and $\rho = 40$ dBm. In Figs.\ \ref{fig_perf_vs_L}(a) and \ref{fig_perf_vs_L}(b), the observations on the achievable rates and EEs of the considered schemes are similar to those from Fig.\ \ref{fig_perf_vs_Pt}. In particular, it is seen that in the entire range of $L$, the proposed ARFA schemes have small losses in total achievable rate but significant improvement in EE with respect to the fixed-activation HBF scheme with $n_l=N$. Furthermore, the proposed ARFA schemes perform better than or comparable to the APS scheme in terms of both achievable rate and EE. In particular, it is clear that the AS is less efficient in both the spectral and energy compared to the proposed schemes. To further explain the EEs, we consider the total power consumption of these schemes in Fig.\ \ref{fig_perf_vs_L}(c). It can be seen that the total power consumption of the fixed-activation schemes quickly increases with $L$. Therefore, activating all $N$ RF chains at all the APs causes an extremely high power consumption for the \cfmimo system, motivating the ARFA in this work. Among the other schemes, the AS scheme consumes the highest power while achieving the lowest rates, making it energy-inefficient, as seen in Fig.\ \ref{fig_perf_vs_L}(b). The proposed {SC-ARFA} and APS schemes have comparable and low power consumption.
	
	To summarize, we can conclude from Figs.\ \ref{fig_perf_vs_Pt} and \ref{fig_perf_vs_L} that when a limited number of RF chains are used, equally activating the same number of RF chains at the APs, as in the fixed-activation HBF scheme with $n_l = N$ or $n_l = \bar{n}$, is relatively energy-inefficient. Furthermore, although the APS scheme is energy-efficient approach, it has losses in total achievable rate. In contrast, the proposed ARFA schemes achieve the highest or nearly highest performance in terms of both spectral and energy efficiency, which are both far better than those of the AS scheme.

	\subsection{Tradeoff between achievable rates and power consumption}

	The total achievable rate and power consumption of the considered schemes versus $\bar{n}$ are evaluated numerically in Fig.\ \ref{fig_perf_vs_nbar} for $L=32$, $K = 8$, $N_r = 64$, $N = 8$, $\bar{n}=\{1,2,\ldots,8\}$, and $\rho = 40$ dBm. From Fig.\ \ref{fig_perf_vs_nbar}, the following observations can be noted:
	\begin{itemize}
		\item {The total achievable rate and power consumption of the fixed-activation HBF  with $n_l=N$ and those of the AS scheme remain unchanged with $\bar{n}$ because $N$ and $N_r^{\text{AS}}$ RF chains, respectively, are always active at every AP.} In contrast, those of the other schemes depend on $\bar{n}$. Specifically, as $\bar{n}$ increases, both the total achievable rate and power consumption of the fixed-activation HBF scheme with $n_l = \bar{n}$, the proposed ARFA, and the APS schemes increase to approach those of the fixed-activation HBF with $n_l=N$.
		
		\item In Fig.\ \ref{fig_rate_vs_nbar}, the proposed ARFA schemes perform closest to the fixed-activation HBF scheme with $n_l=N$, especially for a small $\bar{n}$. In terms of power consumption, they require slightly higher power than the APS scheme. However, the EEs of SC-ARFA and APS are comparable, as shown in Figs.\ \ref{fig_perf_vs_Pt} and \ref{fig_perf_vs_L}, owing to the efficient use of RF chains. Furthermore, for $\bar{n} \ge 2$, the AS scheme has the lowest power consumption at the cost of the smallest total achievable rate.
		
		\item For the optimal performance--power consumption tradeoff in the assumed environment, $\bar{n} \in [2,4]$ can be chosen in the proposed ARFA schemes to achieve a significant power reduction with marginal performance loss. In particular, for $\bar{n}=4$, the performance loss is only $1.4-4.8\%$. With $\bar{n}=1$, only a single RF chain on average is turned on at each AP, and a significant loss in the achievable rate is observed for the proposed ARFA with respect to the fixed-activation HBF with $n_l = N$.
		
	\end{itemize}
	
	\begin{figure}[t]
		\subfigure[Total achievable rate]
		{
			\includegraphics[scale=0.5]{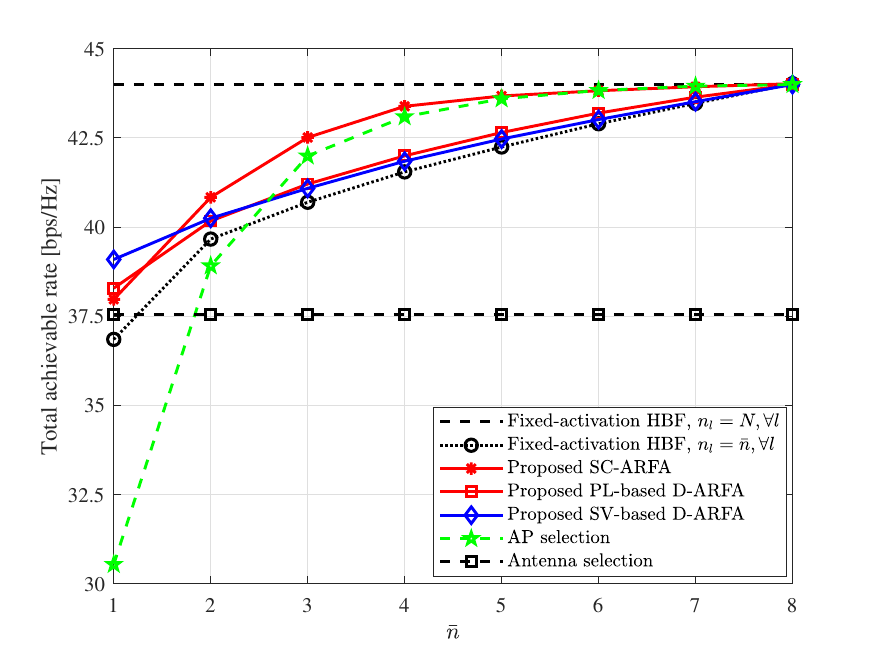}
			\label{fig_rate_vs_nbar}
		}
		\subfigure[Power consumption]
		{
			\includegraphics[scale=0.5]{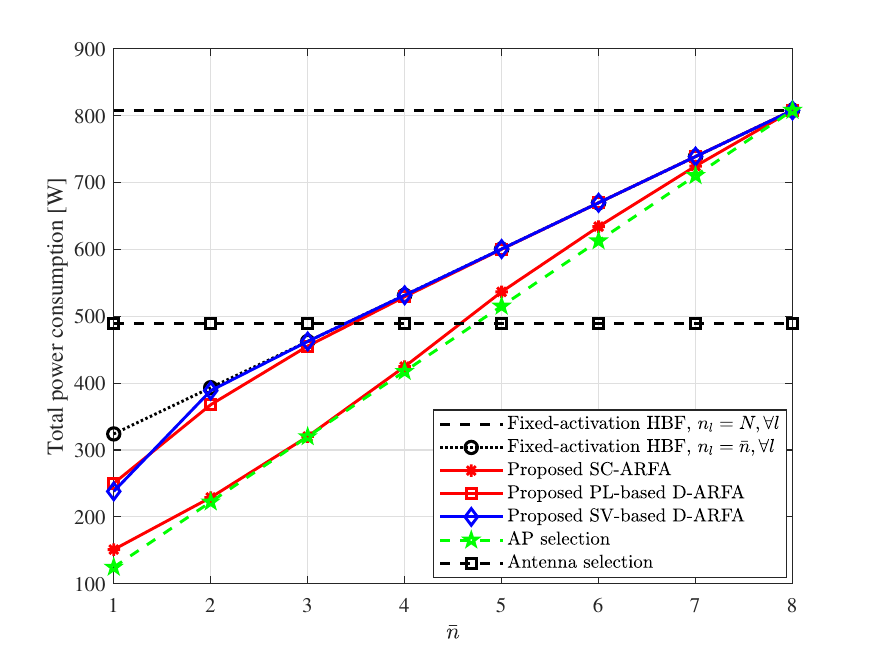}
			\label{fig_power_vs_nbar}
		}
		\caption{Total achievable rates and EEs of the proposed ARFA schemes compared to those of the fixed-activation HBF with $n_l=N$, $n_l=\bar{n}, \forall l$, and APS schemes. Simulation parameters are $L=32$, $K = 8$, $N_r = 64$, $N = 8$, $\bar{n}=\{1,2,\ldots,8\}$, and $\rho = 40$ dBm.}
		\label{fig_perf_vs_nbar}
	\end{figure}

	\subsection{{Fronthauling load analysis}}
	
	\begin{table*}[t]
		\renewcommand{\arraystretch}{0.8}
		\caption{{Comparison of the decentralized and semi-centralized schemes in terms of fronthauling load}}
		\label{tab_info}
		\centering
		\begin{tabular}{|c|c|c|}
			\hline
			
			Schemes   &  AP to CPU &  CPU to AP \\
			\hline
			\hline
			{SC-HBF} & $N_r K$ complex numbers  & $N_r N$ real numbers \\
			\hline
			{SC-ARFA} & $N_r K$ complex numbers  & $N_r \bar{n}$ real numbers  \\
			\hline
			{D-HBF} & {$K$} complex numbers  & 0 \\
			\hline
			{SV-based {D-ARFA}} & {{$K$} complex numbers} and $N$ real numbers  & 1 real number \\
			\hline
			{PL-based {D-ARFA}} & {{$K$} complex numbers}  & {1 real number} \\
			\hline
			
		\end{tabular}
	\end{table*}
	
	In this section, we evaluate the amount of information exchange between the CPU and APs, which is presented in Table \ref{tab_info}. In the {SC-HBF} scheme, the CSI for $\hat{\mH}_l$ of size $N_r \times K$ is sent from the $l$th AP to the CPU, which is used at the CPU to generate $\mF_l$ of size $N_r \times N$. However, we note that all the entries of $\mF_l$ have constant amplitudes of $\frac{1}{\sqrt{N_r}}$. Therefore, only $N_r N$ real numbers representing the phases are fed back on the reverse link. A similar analysis is valid for {SC-ARFA} with the note that only an average of $N_r \bar{n}$ real numbers need to be fed back from the APs to the CPU because in these schemes, only an average of $\bar{n}$ RF chains are activated. It is observed that the amount of information exchange between the CPU and APs is relatively large in the {SC-HBF and SC-ARFA} schemes. 
	
	{In contrast, those for the {decentralized} schemes are small. Specifically, in the {D-HBF} schemes, only $K$ complex numbers representing the estimate of the transmitted signal, i.e., $\vr_l$, are sent to the CPU for the final soft estimation. In the SV-based D-ARFA scheme, an addition of $N$ real numbers for the $N$ singular values are sent from an AP to the CPU for each channel variation to perform ARFA. In contrast, the transmission of path loss values in the PL-based D-ARFA scheme can be ignored because of their slow variations. On the reverse link from the CPU to an AP, only a single real number, which is the number of active RF chains, is fed back in the SV- and PL-based {D-ARFA} schemes, as demonstrated in Algorithms \ref{al_SV_SCARFA} and \ref{al_PL_SCARFA}, respectively. Given that $N \ll N_r$, the decentralized schemes require much less information exchange between the CPU and each AP compared to the semi-centralized schemes.}
	
	\section{Conclusion}
	\label{sec_conclusion}
	
	In this work, we propose two HBF schemes for \cfmimo systems, including {SC-HBF and D-HBF}, in which the analog combiners are generated at the CPU based on the global CSI and at each AP based on the local CSI, respectively. Notably, although the {D-HBF} requires substantially lower computational complexity and no information exchange between the CPU and APs, it achieves approximately the same total achievable rate as that obtained by the {SC-HBF} scheme. Furthermore, to reduce the power consumption in the \cfmimo system, we propose adaptive activation of RF chains at the APs. Low-complexity algorithms are developed to select the number of active RF chains at the APs such that the system's power consumption is significantly reduced with only a marginal loss in the total achievable rate. The efficiency of the proposed schemes is justified by the simulation results, which show that the proposed ARFA scheme achieves significant improvement in EE while leading to a loss of only small loss in total achievable rate.  In this work, the assumption on the availability of the full CSI for all the antennas is adopted for the design of hybrid beamformers and ARFA schemes. Although such full CSI can be obtained by compressed sensing-based approaches, further studies are required to make it practical in CF mmWave massive MIMO systems. Further optimization of the proposed analog combining schemes for wideband systems can be considered for future research. Furthermore, the proposed ARFA scheme can be incorporated with low-resolution ADCs \cite{zhang2020prospective, zhang2018low} to further reduce the power consumption.
	
	\appendices
	\section{Proof of Theorem \ref{theorem_rate_C}}
	\label{appendix_rate_C}
	Let $\mQ = \mI_{LN} + \gamma \mF^H \mH \mH^H \mF$. Because $\mF$ is a block-diagonal matrix, we have $$\mH^H \mF = \left[\mH_1^H \mF_1, \mH_2^H \mF_2, \ldots, \mH_L^H \mF_L \right],$$
	leading to $\mH^H \mF \mF^H \mH =  \sum_{l=1}^{L} \mH_l^H \mF_l \mF_l^H \mH_l$. Therefore, $\mQ$ can be expressed as $\mQ = \mI_{K}+  \gamma \sum_{l=1}^{L} \mH_l^H \mF_l \mF_l^H \mH_l$. By letting $\mG_l = \mH_l^H \mF_l \mF_l^H \mH_l$, $\mQ$ can be further expanded as
	\begin{align*}
	\mQ &= \underbrace{\mI_{K}+  \gamma \mG_1}_{\triangleq \mE_1} + \gamma \mG_2 + \ldots + \gamma \mG_L  = \mE_1 (\underbrace{\mI_{K} + \gamma \mE_1^{-1} \mG_2}_{\triangleq \mE_2} + \ldots + \gamma \mE_1^{-1} \mG_L) \\
	&= \mE_1 \mE_2  (\underbrace{\mI_{K} + \gamma \mE_2^{-1} \mE_1^{-1} \mG_2}_{\triangleq \mE_3} + \ldots + \gamma \mE_2^{-1} \mE_1^{-1} \mG_L) = \ldots = \mE_1 \mE_2 \ldots \mE_L, \numberthis \label{eqn_Q}
	\end{align*}
	where $\mE_l = \mI_{K} + \gamma (\mE_1 \ldots \mE_{l-1})^{-1} \mG_l$, $l=2,3,\ldots,L$. As a result, {$\bar{\mathcal{R}}^{\mathrm{a}}$} can be expressed as
	\begin{align*}
	{\bar{\mathcal{R}}^{\mathrm{a}}} &= \log_2 \det \mQ  = \sum_{l=1}^{L} \log_2 \det (\mE_l) = \sum_{l=1}^{L} \log_2 \det (\mI_{K} + \gamma (\underbrace{\mE_1 \ldots \mE_{l-1}}_{\triangleq \mQ_{l-1}})^{-1} \mG_l) \numberthis \label{def_Ql1} \\
	&=  \sum_{l=1}^{L} \log_2 \det \left(\mI_{K} + \gamma \mQ_{l-1}^{-1} \mH_l^H \mF_l \mF_l^H \mH_l \right) =  \sum_{l=1}^{L} \log_2 \det \left(\mI_{N} + \gamma \mF_l^H \mH_l \mQ_{l-1}^{-1} \mH_l^H \mF_l \right), \numberthis \label{R_sum}
	\end{align*}
	as given in Theorem \ref{theorem_rate_C}. The last equality in \eqref{R_sum} follows from $\det(\mI_{K} + \mA \mB) = \det(\mI_{N} + \mB \mA)$ with $\mA = \mQ_{l-1}^{-1} \mH_l^H \mF_l \in \setC^{K \times N}$ and $\mB = \mF_l^H \mH_l  \in \setC^{N \times K}$.  Furthermore, from the definition of $\mQ_{l-1}$ in \eqref{def_Ql1}, we have $\mE_{l-1} = \mI_{K} + \gamma (\mE_1 \ldots \mE_{l-2})^{-1} \mG_{l-1} = \mI_{K} + \gamma \mQ_{l-2}^{-1} \mG_l$. Finally, recalling that $\mG_l = \mH_l^H \mF_l \mF_l^H \mH_l$, we obtain the expression of $\mQ_{l-1}$ in \eqref{Q_n_1}, i.e.,
	\begin{align*}
	\mQ_{l-1} &= \mQ_{l-2} \mE_{l-1}  = \mQ_{l-2} (\mI_{K} + \gamma \mQ_{l-2}^{-1} \mG_l)  
	= \mQ_{l-2} + \gamma \mH_{l-1}^H \mF_{l-1} \mF_{l-1}^H \mH_{l-1}, \numberthis \label{Q_l1}
	\end{align*}
	with $\mQ_0 = \mI_{K}$, which completes the proof.
	
	\section{Proof of Remark 1}
	\label{appendix_rate_SC}
	
	From \eqref{Q_l1}, $\mQ_{l-1}$ in \eqref{R_n} can be expressed as
	\begin{align*}
	\mQ_{l-1} = \mI_{K}+  \gamma \sum_{i=1}^{l-1} \mH_i^H \mF_i \mF_i^H \mH_i, l=2,\ldots,L. \numberthis \label{Q_l_diag}
	\end{align*}
	
	\subsubsection{When $l$ is small}
	With the assumption of very low SNRs in CF mmWave massive MIMO, we have $\mQ_{l-1} \approx \mI_{K}$ for small $l$, leading to
	\begin{align*}
	\bar{\mathcal{R}}^{\mathrm{a}}_l \approx \tilde{\mathcal{R}}^{\mathrm{a}}_l = \log_2 \det \left( \mI_{N} + \gamma \mF_l^H \mH_l \mH_l^H \mF_l \right), \numberthis \label{R_small_l}
	\end{align*}
	where $\bar{\mathcal{R}}^{\mathrm{a}}_l$ is the sub-rate associated with the $l$th AP in {SC-HBF}, given in \eqref{R_n}. The unconstrained combiner that maximizes $\tilde{\mathcal{R}}_l$ in \eqref{R_small_l} is the matrix with columns being the $N$ singular vectors corresponding to the $N$ largest singular values of $\mH_l$. As a result, the analog combining vectors in the {D-HBF} scheme can be determined as in \eqref{f_tilde}.
	
	\subsubsection{As $l$ increases}
	Because $\mF_i$ only depends on $\mH_i$ for small $l$, $\{\mH_i^H \mF_i \mF_i^H \mH_i\}, i=1, \ldots, l-1$ are independent of each other. As $l$ grows and becomes sufficiently large, by the law of large numbers,  we have $\sum_{i=1}^{l-1} \mH_i^H \mF_i \mF_i^H \mH_i \rightarrow (l-1) \bar{\mE}$, where $\bar{\mE}= \mean{\mH_i^H \mF_i \mF_i^H \mH_i}$ has constant diagonal elements and zeros for off-diagonal elements. Therefore, $\mQ_{l-1}$ in \eqref{Q_l_diag} becomes approximately diagonal even when $l$ is large, as does $\mQ_{l-1}^{-1}$.
	
	Based on the ordered singular value decomposition, $\mH_l$ can be factorized as $\mH_l = \mU \mathbf{\Sigma} \mV^H$, where $\mathbf{\Sigma}$ is an $N_r \times K$ rectangular diagonal matrix with the singular values of $\mH_l$ on the main diagonal in decreasing order, whereas $\mU$ and $\mV$ are unitary matrices of size $N_r \times N_r$ and $K \times K$, whose columns are the left- and right-singular vectors of $\mH_l$, respectively. Then, $\bar{\mathcal{R}}_l^{\mathrm{a}}$ in \eqref{R_n} can be expressed as
	\begin{align*}
	\bar{\mathcal{R}}_l^{\mathrm{a}} &= \log_2 \det ( \mI_{N} + \gamma \mF_l^H \mU \underbrace{\mathbf{\Sigma} \mV^H \mQ_{l-1}^{-1} \mV \mathbf{\Sigma}^H}_{\triangleq \mathbf{\Lambda}} \mU^H \mF_l ).  \numberthis \label{T_1}
	\end{align*} 
	Because $\mQ_{l-1}^{-1}$ is approximately a diagonal matrix with constant diagonal elements, as shown above, $\mathbf{\Lambda} = \mathbf{\Sigma} \mV^H \mQ_{l-1}^{-1} \mV \mathbf{\Sigma}^H$ becomes approximately a diagonal matrix, and its diagonal elements are in decreasing order. Therefore, the optimal solution of $\max_{\mF_l} \bar{\mathcal{R}}_l^{\mathrm{a}} = \log_2 \det ( \mI_{N} + \gamma \mF_l^H \mU \mathbf{\Lambda} \mU^H \mF_l )$	is approximately the matrix with columns being the first $N$ columns of $\mU$, which are the singular vectors corresponding to the $N$ largest singular values of $\mH_l$, implying the analog combining vectors given in \eqref{f_tilde} for {D-HBF}. This completes the proof.

	\bibliographystyle{IEEEtran}
	\bibliography{IEEEabrv,Bibliography}

\end{document}